\documentclass[aps,twocolumn,showpacs,preprintnumbers,amsmath,amssymb,nofootinbib,showkeys,floatfix]{revtex4-1}

\usepackage{epsfig}
\usepackage{graphicx}
\usepackage{dcolumn}

\usepackage{latexsym}
\usepackage{graphicx}
\usepackage{enumerate}
\usepackage{float}
\usepackage{placeins}
\usepackage{supertabular}

\def\tstruta{\vrule height3.0ex depth0pt width0pt} 
\def\tstrutb{\vrule height2.5ex depth0pt width0pt} 

\newcommand{\bq}{$b\,$-}

\newcommand{\req}[1]{Eq.~(\ref{#1})}

\begin{document}

\title{Semileptonic $B$ and $B_s$ decays into orbitally excited charmed mesons}
\author{J. Segovia}
\author{C. Albertus}
\author{D.R. Entem}
\author{F. Fern\'andez}
\author{E. Hern\'andez}
\author{M.A. P\'erez-Garc\'ia}
\affiliation{Departamento de F\'isica Fundamental e IUFFyM, \\ Universidad de 
Salamanca, E-37008 Salamanca, Spain}
\date{\today}

\begin{abstract}

The BaBar Collaboration has recently reported products of branching fractions 
that include $B$ meson semileptonic decays  into final states with charged and 
neutral $D_{1}(2420)$ and $D_{2}^{\ast}(2460)$, two narrow orbitally excited
charmed mesons. We evaluate these branching fractions, together with those
concerning $D_{0}^{\ast}(2400)$ and $D_{1}'(2430)$ mesons, within the framework
of a constituent quark model. The calculation is performed in two steps, one of
which involves a semileptonic decay and the other is mediated by a strong
process. Our results are in agreement with the experimental data. We also extend
the study to semileptonic decays of $B_{s}$ into orbitally excited
charmed-strange mesons, providing predictions to the possible measurements to be
carried out at LHC.

\end{abstract}


\maketitle

\section{INTRODUCTION}
\label{sec:introduction}

Different collaborations have recently reported semileptonic $B$ decays into
orbitally excited charmed mesons providing detailed results of branching
fractions. The theoretical analysis of these data, which include both weak and
strong decays, offers the possibility for a stringent test of meson models. 

Moreover, an accurate determination of the $|V_{cb}|$ and $|V_{ub}|$
Cabbibo\,-\,Kobayashi\,-\,Maskawa matrix elements demands a detailed knowledge
of semileptonic decays of \bq hadrons. Decays including orbitally excited
charmed meson in the final state provide a substantial contribution to the total
semileptonic decay width. Furthermore, a better understanding of these processes
is also necessary in the analysis of signals and backgrounds of inclusive and
exclusive measurements of \bq hadron decays.

The Belle Collaboration~\cite{belle}, using a full reconstruction tagging method
to suppress the large combinatorial background, reported data on the product of
branching fractions $\mathcal{B}(B^{+} \to
D^{\ast\ast}l^{+}\nu_{l})\mathcal{B}(D^{\ast\ast} \to D^{(\ast)}\pi)$, where,
in the usual notation, $l$ stands for a light $e$ or $\mu$ lepton, the
$D_{0}^{\ast}$, $D_{1}'$, $D_{1}$ and $D_{2}^{\ast}$ mesons are denoted
generically as $D^{\ast\ast}$, and the $D^{\ast}$ and $D$ mesons as
$D^{(\ast)}$.

$D^{\ast\ast}$ decays are reconstructed in the decay chains $D^{\ast\ast} \to
D^{\ast}\pi^{\pm}$ and $D^{\ast\ast} \to D\pi^{\pm}$. In particular, the
$D_{0}^{\ast}$ meson decays only through the $D\pi$ channel, while the $D_{1}'$
and $D_{1}$ mesons decay only via $D^{\ast}\pi$. Both $D\pi$ and $D^{\ast}\pi$
channels are opened  in the case of $D_{2}^{\ast}$.

In the case of BaBar data~\cite{aubert08,aubert09} the branching fractions
$\mathcal{B}(D_{2}^{\ast} \to D^{(\ast)}\pi)$ include both the $D^{\ast}$ and
$D$ contributions. As they also provide the ratio $\mathcal{B}_{D/D^{(\ast)}}$
we  estimate  the $D^{\ast}$ and $D$ contributions separately. The experimental
results of both collaborations are given in Table~\ref{tab:experiment}. 

\begin{table*}[t!]
\begin{center}
\begin{tabular}{lcc}
\hline
\hline
\tstrutb
& Belle~\cite{belle} $(\times10^{-3})$ & BaBar~\cite{aubert08,aubert09}
$(\times10^{-3})$ \\
\hline
\tstrutb
$D_{0}^{\ast}(2400)$ & & \\[2ex]
${\cal B}(B^+ \to \bar{D}^{\ast0}_{0} l^+ \nu_l){\cal B}(\bar{D}^{\ast0}_{0}\to
D^{-}\pi^+)$ & $2.4\pm0.4\pm0.6$ & $2.6\pm0.5\pm0.4$ \\
${\cal B}(B^0 \to D^{\ast-}_{0} l^+ \nu_l){\cal B}(D^{\ast-}_{0}\to
\bar{D}^{0}\pi^-)$ & $2.0\pm0.7\pm0.5$ & $4.4\pm0.8\pm0.6$ \\
\hline
\tstrutb
$D'_{1}(2430)$ & & \\[2ex]
${\cal B}(B^+ \to \bar{D}^{'0}_{1} l^+ \nu_l){\cal B}(\bar{D}^{'0}_{1}\to
D^{\ast-}\pi^+)$ & $<0.7$ & $2.7\pm0.4\pm0.5$ \\
${\cal B}(B^0 \to D^{'-}_{1} l^+ \nu_l){\cal B}(D^{'-}_1\to
\bar{D}^{\ast0}\pi^-)$ & $<5$ & $3.1\pm0.7\pm0.5$ \\
\hline
\tstrutb
$D_{1}(2420)$ & & \\[2ex]
${\cal B}(B^+ \to \bar{D}^0_1 l^+ \nu_l){\cal B}(\bar{D}^0_1\to D^{\ast-}\pi^+)$
& $4.2\pm0.7\pm0.7$ & $2.97\pm0.17\pm0.17$ \\
${\cal B}(B^0 \to D^-_1 l^+ \nu_l){\cal B}(D^-_1\to \bar{D}^{\ast0}\pi^-)$ &
$5.4\pm1.9\pm0.9$ & $2.78\pm0.24\pm0.25$ \\
\hline
\tstrutb
$D_{2}^{\ast}(2460)$ & & \\[2ex]
${\cal B}(B^+ \to \bar{D}^{\ast0}_2 l^+ \nu_l){\cal B}(\bar{D}^{\ast0}_2\to
D^{-}\pi^+)$ & $2.2\pm0.3\pm0.4$ & $1.4\pm0.2\pm0.2^{(\ast)}$ \\
${\cal B}(B^+ \to \bar{D}^{\ast0}_2 l^+ \nu_l){\cal B}(\bar{D}^{\ast0}_2\to
D^{\ast-}\pi^+)$ & $1.8\pm0.6\pm0.3$ & $0.9\pm0.2\pm0.2^{(\ast)}$ \\
${\cal B}(B^+ \to \bar{D}^{\ast0}_2 l^+ \nu_l){\cal B}(\bar{D}^{\ast0}_2\to
D^{(\ast)-}\pi^+)$ & $4.0\pm0.7\pm0.5$ & $2.3\pm0.2\pm0.2$ \\[2ex]
${\cal B}(B^0 \to D^{\ast-}_2 l^+ \nu_l){\cal B}(D^{\ast-}_2\to
\bar{D}^{0}\pi^-)$ & $2.2\pm0.4\pm0.4$ & $1.1\pm0.2\pm0.1^{(\ast)}$ \\
${\cal B}(B^0 \to D^{\ast-}_2 l^+ \nu_l){\cal B}(D^{\ast-}_2\to
\bar{D}^{\ast0}\pi^-)$ & $<3$ & $0.7\pm0.2\pm0.1^{(\ast)}$ \\
${\cal B}(B^0 \to D^{\ast-}_2 l^+ \nu_l){\cal B}(D^{\ast-}_2\to
\bar{D}^{(\ast)0}\pi^-)$ & $<5.2$ & $1.8\pm0.3\pm0.1$ \\[2ex]
${\cal B}_{D/D^{(\ast)}}$ & $0.55\pm0.03$ & $0.62\pm0.03\pm0.02$ \\
\hline
\hline
\end{tabular}
\caption{\label{tab:experiment} Most recent experimental measurements reported
by Belle and BaBar Collaborations. $l$ stands for a light $e$ or $\mu$
lepton. The symbol ${(\ast)}$ indicates  results estimated from the original data by using
$B_{D/D^{(\ast)}}$.}
\end{center}
\end{table*}

A similar analysis can be done in the strange sector for the $B_{s}$ meson
semileptonic decays. Here the intermediate states are the orbitally
charmed-strange mesons, $D_{s}^{\ast\ast}$, and the available final channels are
$DK$ and $D^{\ast}K$. The Particle Data Group (PDG) reports a value
 ${\cal B}(B_{s}^{0} \to D_{s1}(2536)^{-} \mu^+
\nu_\mu){\cal B}(D_{s1}(2536)^{-}\to
D^{\ast-}\bar{K}^{0})=2.4\pm0.7$~\cite{PDG2010} based on 
their best value for ${\cal B}(\bar b\to B_s^0)$ and the experimental data 
for ${\cal B}(\bar b\to B_s^0){\cal B}(B_{s}^{0} \to D_{s1}(2536)^{-} \mu^+
\nu_\mu){\cal B}(D_{s1}(2536)^{-}\to
D^{\ast-}\bar{K}^{0})$ measured by the D0 
Collaboration~\cite{abazov09}.

All these magnitudes can be consistently calculated in the framework of
constituent quark models because they can simultaneously account for the
hadronic part of the weak process and the strong meson decays. In this context,
meson strong decay has been described successfully in phenomenological models,
like the $^{3}P_{0}$ model~\cite{mic69_1} or the flux-tube model~\cite{kok87_1},
or in microscopic models (see Refs.~\cite{Eichten78,Swanson96}). The difference
between the two approaches lies on the description of the $q\bar{q}$ creation
vertex. While the $^{3}P_{0}$ model assumes that the $q\bar{q}$ pair is created
from the vacuum with vacuum quantum numbers, in the microscopic model the
$q\bar{q}$ pair is created from the  interquark interactions already acting in
the model. Both approaches will be used here to evaluate the strong decays.
As for the weak process the matrix elements  factorizes into a leptonic and 
a hadronic part. It is the hadronic part that contains the nonperturbative 
strong interaction effects and we shall evaluate it within a constituent quark
model (CQM). We will work within the CQM of Ref.~\cite{Vijande2005} which
successfully describes hadron phenomenology and hadronic
reactions~\cite{Fernandez1992,Garcilazo2001,Vijande2001} and has recently been
applied to mesons containing heavy quarks in
Refs.~\cite{Segovia2008,Segovia2009}.

The paper is organized as follows: In Sec.~\ref{sec:CQM} we introduce the
model we have used to get the masses and wave functions of the mesons involved
in the reactions mentioned above. In Secs.~\ref{sec:weakdecays} and
\ref{sec:strongdecays} we study the semileptonic and strong decay
mechanisms, which constitute the two steps of the processes under study.
Finally, we present our results in Sec.~\ref{sec:results} and give some
conclusions in Sec.~\ref{sec:conclusions}.

\section{CONSTITUENT QUARK MODEL}
\label{sec:CQM}

Spontaneous chiral symmetry breaking of the QCD Lagrangian together with the
perturbative one-gluon exchange (OGE) and the nonperturbative confining
interaction are the main pieces of potential models. Using this idea, Vijande
{\it et al.}~\cite{Vijande2005} developed a model of the quark-quark interaction
which is able to describe meson phenomenology from the light to the heavy quark
sector. We briefly explain the model below. Further details can be found in
Ref.~\cite{Vijande2005}. 

One consequence of the spontaneous chiral symmetry breaking is that the nearly
massless 'current' light quarks acquire a dynamical, momentum-dependent mass
$M(p)$ with $M(0)\approx 300\,\mbox{MeV}$ for the $u$ and $d$ quarks, namely,
the constituent mass. To preserve chiral invariance of the QCD Lagrangian new
interaction terms, given by Goldstone boson exchanges, should appear between
constituent quarks.

A simple Lagrangian invariant under chiral transformations can be derived
as~\cite{diakonov2003}
\begin{eqnarray}
{\mathcal L} &=& \bar \psi ( i\gamma^\mu \partial_\mu - M U^{\gamma_5})\psi,
\end{eqnarray}
where $U^{\gamma_5}=\exp(i\pi^a\lambda^a\gamma_5/f_\pi)$, $\pi^a$ denotes the
pseudoscalar fields $(\vec \pi,K,\eta_8)$ and $M$ is the constituent quark mass.
The momentum-dependent mass acts as a natural cutoff of the theory. The chiral
quark-quark interaction can be written as
\begin{align}
V_{qq}\left(\vec{r}_{ij}\right)=V_{qq}^{\rm C}\left(\vec{r}_{ij}\right)
+V_{qq}^{\rm T}\left(\vec{r}_{ij}\right)+V_{qq}^{\rm
SO}\left(\vec{r}_{ij}\right),
\end{align}
where $C$, $T$ and $SO$ stand for central, tensor and spin-orbit potentials.
The central part presents four different contributions,
\begin{equation}
V_{qq}^{\rm C}\left(\vec{r}_{ij}\right)=V_{\pi}^{\rm C}\left(\vec{r}_{ij}\right)
+V_{\sigma}^{\rm C}\left(\vec{r}_{ij}\right)+V_{K}^{\rm
C}\left(\vec{r}_{ij}\right)+V_{\eta}^{\rm C}\left(\vec{r}_{ij}\right),
\end{equation}
given by
\begin{widetext}
\begin{equation}
\begin{split}
&
V_{\pi}^{\rm C}\left( \vec{r}_{ij} \right)= \frac{g_{ch}^{2}}{4\pi}
\frac{m_{\pi}^2}{12m_{i}m_{j}}\frac{\Lambda_{\pi}^{2}}{\Lambda_{\pi}^{2}-m_{\pi}
^{2}}m_{\pi}\left[Y(m_{\pi}r_{ij})-\frac{\Lambda_{\pi}^{3}}{m_{\pi}^{3}}
Y(\Lambda_{\pi}r_{ij}) \right]
(\vec{\sigma}_{i}\cdot\vec{\sigma}_{j})\sum_{a=1}^{3}(\lambda_{i}^{a}
\cdot\lambda_{j}^{a}), \\
& 
V_{\sigma}^{\rm C}\left( \vec{r}_{ij} \right) =-\frac{g_{ch}^{2}}{4\pi}
\frac{\Lambda_{\sigma}^{2}}{\Lambda_{\sigma}^{2}-m_{\sigma}^{2}}m_{\sigma}\left[
Y(m_{\sigma}r_{ij})-\frac{\Lambda_{\sigma}}{m_{\sigma}}Y(\Lambda_{\sigma}r_{ij})
\right], \\
& 
V_{K}^{\rm C}\left( \vec{r}_{ij} \right)= \frac{g_{ch}^{2}}{4\pi}
\frac{m_{K}^2}{12m_{i}m_{j}}\frac{\Lambda_{K}^{2}}{\Lambda_{K}^{2}-m_{K}^{2}}m_{
K}\left[Y(m_{K}r_{ij})-\frac{\Lambda_{K}^{3}}{m_{K}^{3}}Y(\Lambda_{K}r_{ij})
\right] (\vec{\sigma}_{i}\cdot\vec{\sigma}_{j})\sum_{a=4}^{7}(\lambda_{i}^{a}
\cdot\lambda_{j}^{a}), \\
& 
V_{\eta}^{\rm C}\left( \vec{r}_{ij} \right)= \frac{g_{ch}^{2}}{4\pi}
\frac{m_{\eta}^2}{12m_{i}m_{j}}\frac{\Lambda_{\eta}^{2}}{\Lambda_{\eta}^{2}-m_{
\eta}^{2}}m_{\eta}\left[Y(m_{\eta}r_{ij})-\frac{\Lambda_{\eta}^{3}}{m_{\eta}^{3}
}Y(\Lambda_{\eta}r_{ij}) \right] (\vec{\sigma}_{i}\cdot\vec{\sigma}_{j})
\left[\cos\theta_{p}\left(\lambda_{i}^{8}\cdot\lambda_{j}^{8}
\right)-\sin\theta_{p}\right],
\end{split}
\end{equation}
\end{widetext}
where $Y(x)$ is the standard Yukawa function defined by $Y(x)=e^{-x}/x$. We
consider the physical $\eta$ meson instead of the octet one and so we introduce
the angle $\theta_p$. The $\lambda^{a}$ are the SU(3) flavor Gell-Mann
matrices, $m_{i}$ is the quark mass and $m_{\pi}$, $m_{K}$ and $m_{\eta}$ are 
the masses of the SU(3) Goldstone bosons, taken from experimental values.
$m_{\sigma}$ is determined through the PCAC relation $m_{\sigma}^{2}\simeq
m_{\pi}^{2}+4m_{u,d}^{2}$~\cite{sca82_1}. Finally, the chiral coupling
constant, $g_{ch}$, is determined from the $\pi NN$ coupling constant through
\begin{equation}
\frac{g_{ch}^{2}}{4\pi}=\frac{9}{25}\frac{g_{\pi NN}^{2}}{4\pi}
\frac{m_{u,d}^{2}}{m_{N}^2},
\end{equation}
which assumes that flavor SU(3) is an exact symmetry only broken by the
different mass of the strange quark.

There are three different contributions to the tensor potential
\begin{equation}
V_{qq}^{\rm T}(\vec{r}_{ij}) = V_{\pi}^{\rm T}(\vec{r}_{ij}) +
V_{K}^{\rm T}(\vec{r}_{ij}) + V_{\eta}^{\rm T}(\vec{r}_{ij}),
\end{equation}
given by
\begin{widetext}
\begin{equation}
\begin{split}
&
V_{\pi}^{\rm T}\left( \vec{r}_{ij} \right)= \frac{g_{ch}^{2}}{4\pi}
\frac{m_{\pi}^2}{12m_{i}m_{j}}\frac{\Lambda_{\pi}^{2}}{\Lambda_{\pi}^{2}-m_{\pi}
^{2}}m_{\pi}\left[H(m_{\pi}r_{ij})-\frac{\Lambda_{\pi}^{3}}{m_{\pi}^{3}}
H(\Lambda_{\pi}r_{ij}) \right] S_{ij}
\sum_{a=1}^{3}(\lambda_{i}^{a}\cdot\lambda_{j}^{a}), \\
& 
V_{K}^{\rm T}\left( \vec{r}_{ij} \right)= \frac{g_{ch}^{2}}{4\pi}
\frac{m_{K}^2}{12m_{i}m_{j}}\frac{\Lambda_{K}^{2}}{\Lambda_{K}^{2}-m_{K}^{2}}m_{
K}\left[H(m_{K}r_{ij})-\frac{\Lambda_{K}^{3}}{m_{K}^{3}}H(\Lambda_{K}r_{ij})
\right] S_{ij} \sum_{a=4}^{7}(\lambda_{i}^{a}\cdot\lambda_{j}^{a}), \\
& 
V_{\eta}^{\rm T}\left( \vec{r}_{ij} \right)= \frac{g_{ch}^{2}}{4\pi}
\frac{m_{\eta}^2}{12m_{i}m_{j}}\frac{\Lambda_{\eta}^{2}}{ \Lambda_{\eta}^{2}-m_{
\eta}^{2}}m_{\eta}\left[H(m_{\eta}r_{ij})-\frac{\Lambda_{\eta}^{3}}{m_{\eta}^{3}
}H(\Lambda_{\eta}r_{ij}) \right] S_{ij} \left[ \cos \theta_{p} \left(
\lambda_{i}^{8}\cdot\lambda_{j}^{8} \right)-\sin\theta_{p}\right].
\end{split}
\end{equation}
\end{widetext}
$S_{ij}=3(\vec{\sigma}_{i}\cdot\hat{r}_{ij})(\vec{\sigma}_{j}\cdot\hat{r}_{ij})
- \vec{\sigma_{i}}\cdot\vec{\sigma_{j}}$ is the quark tensor operator and
$H(x)=(1+3/x+3/x^{2})Y(x)$.

Finally, the spin-orbit potential only presents a contribution coming from 
the scalar part of the interaction
\begin{align}
&
V_{qq}^{\rm SO}(\vec{r}_{ij})=V_{\sigma}^{\rm SO}\left( \vec{r}_{ij} \right) = 
-\frac{g_{ch}^{2}}{4\pi}\frac{m_{\sigma}^{3}}{2m_{i}m_{j}}\frac{\Lambda_{\sigma}
^{2}}{\Lambda_{\sigma}^{2}-m_{\sigma}^{2}} \nonumber \\ 
&
\times\left[G(m_{\sigma}r_{ij})-\frac{\Lambda_{\sigma}^{3}}{m_{\sigma}^{3}}
G(\Lambda_{\sigma}r_{ij}) \right] (\vec{L}\cdot\vec{S}).
\end{align}
In the last equation $G(x)$ is the function $(1+1/x)Y(x)/x$.

Beyond the chiral symmetry breaking scale one expects the dynamics to be
governed by QCD perturbative effects. In this way one-gluon fluctuations around
the instanton vacuum are taken into account through the $qqg$ coupling
\begin{eqnarray}
{\mathcal L}_{qqg} &=& i\sqrt{4\pi\alpha_s} \bar \psi \gamma_\mu G^\mu_c
\lambda^c \psi,
\label{Lqqg}
\end{eqnarray}
with $\lambda^c$ being the $SU(3)$ color matrices and $G^\mu_c$ the gluon field.

The different terms of the potential derived from the Lagrangian contain
central, tensor, and spin-orbit contributions and are given by
\begin{widetext}
\begin{align}
V_{\rm OGE}^{\rm C}(\vec{r}_{ij})=
&\frac{1}{4}\alpha_{s}(\vec{\lambda}_{i}^{c}\cdot
\vec{\lambda}_{j}^{c})\left[ \frac{1}{r_{ij}}-\frac{1}{6m_{i}m_{j}} 
(\vec{\sigma}_{i}\cdot\vec{\sigma}_{j}) 
\frac{e^{-r_{ij}/r_{0}(\mu)}}{r_{ij}r_{0}^{2}(\mu)}\right], \nonumber \\
V_{\rm OGE}^{\rm T}(\vec{r}_{ij})= &-\frac{1}{16}\frac{\alpha_{s}}{m_{i}m_{j}}
(\vec{\lambda}_{i}^{c}\cdot\vec{\lambda}_{j}^{c})\left[ 
\frac{1}{r_{ij}^{3}}-\frac{e^{-r_{ij}/r_{g}(\mu)}}{r_{ij}}\left( 
\frac{1}{r_{ij}^{2}}+\frac{1}{3r_{g}^{2}(\mu)}+\frac{1}{r_{ij}r_{g}(\mu)}\right)
\right]S_{ij}, \nonumber \\
 V_{\rm OGE}^{\rm SO}(\vec{r}_{ij}) = &
-\frac{1}{16}\frac{\alpha_{s}}{m_{i}^{2}m_{j}^{2}}(\vec{\lambda}_{i}^{c}\cdot
\vec{\lambda}_{j}^{c})\left[\frac{1}{r_{ij}^{3}}-\frac{e^{-r_{ij}/r_{g}(\mu)}}{
r_{ij}^{3}} \left(1+\frac{r_{ij}}{r_{g}(\mu)}\right)\right] \times \nonumber \\
&
\times \left[((m_{i}+m_{j})^{2}+2m_{i}m_{j})(\vec{S}_{+}\cdot\vec{L})+(m_{j}^{2}
-m_{i}^{2}) (\vec{S}_{-}\cdot\vec{L}) \right],
\end{align}
\end{widetext}
where $\vec{S}_{\pm}=\frac{1}{2}(\vec{\sigma}_{i}\,\pm\,\vec{\sigma}_{j})$.
Besides, 
$r_{0}(\mu)=\hat{r}_{0}\frac{\mu_{nn}}{\mu_{ij}}$ and
$r_{g}(\mu)=\hat{r}_{g}\frac{\mu_{nn}}{\mu_{ij}}$ are regulators which depend on
$\mu_{ij}$, the reduced mass of the $q\bar{q}$ pair. The contact term of the
central potential has been regularized as
\begin{equation}
\delta(\vec{r}_{ij})\sim\frac{1}{4\pi r_{0}^{2}}\frac{e^{-r_{ij}/r_{0}}}{r_{ij}}
\end{equation}

The wide energy range needed to provide a consistent description of light,
strange and heavy mesons requires an effective scale-dependent strong coupling
constant. We use the frozen coupling constant of Ref.~\cite{Vijande2005}
\begin{equation}
\alpha_{s}(\mu)=\frac{\alpha_{0}}{\ln\left( 
\frac{\mu^{2}+\mu_{0}^{2}}{\Lambda_{0}^{2}} \right)},
\end{equation}
in which $\mu$ is the reduced mass of the $q\bar{q}$ pair and $\alpha_{0}$,
$\mu_{0}$ and $\Lambda_{0}$ are parameters of the model determined by a global
fit to the meson spectra.

Confinement is one of the crucial aspects of QCD. Color charges are confined
inside hadrons. It is well known that multigluon exchanges produce an attractive
linearly rising potential proportional to the distance between quarks. This idea
has been confirmed, but not rigorously proved, by quenched lattice gauge Wilson
loop calculations for heavy valence quark systems. However, sea quarks are also
important ingredients of the strong interaction dynamics. When included in the
lattice calculations they contribute to the screening of the rising potential at
low momenta and eventually to the breaking of the quark-antiquark binding
string. This fact, which has been observed in $n_f=2$ lattice
QCD~\cite{bali2005}, has been taken into account in our model by including the
terms
\begin{widetext}
\begin{align}
V_{\rm CON}^{\rm C}(\vec{r}_{ij}) =& \left[ -a_{c}(1-e^{-\mu_{c}r_{ij}})+\Delta
\right] (\vec{\lambda}_{i}^{c}\cdot\vec{\lambda}_{j}^{c}), \nonumber\\
V_{\rm CON}^{\rm SO}(\vec{r}_{ij})= &
-\left(\vec{\lambda}_{i}^{c}\cdot\vec{\lambda}_{j}^{c} \right) 
\frac{a_{c}\mu_{c}e^{-\mu_{c}r_{ij}}}{4m_{i}^{2}m_{j}^{2}r_{ij}}\left[((m_{i}^{2
}+m_{j}^{2})(1-2a_{s}) +4m_{i}m_{j}(1-a_{s}))(\vec{S}_{+}\cdot\vec{L})\right.
\nonumber \\
&
\left. +(m_{j}^{2}-m_{i}^{2})(1-2a_{s})(\vec{S}_{-}\cdot\vec{L}) \right], 
\end{align}
\end{widetext}
where $a_{s}$ controls the mixture between the scalar and vector Lorentz
structures of the confinement. At short distances this potential presents a
linear behavior with an effective confinement strength
$\sigma=-a_{c}\,\mu_{c}\,(\vec{\lambda}^{c}_{i} \cdot \vec{\lambda}^{c}_{j})$
and becomes constant at large distances with a threshold defined by
\begin{equation}
V_{\rm thr}=\{-a_{c}+ \Delta\}(\vec{\lambda}^{c}_{i}\cdot
\vec{\lambda}^{c}_{j}).
\label{eq:threshold}
\end{equation}

No $q\bar{q}$ bound states can be found for energies higher than this threshold.
The system suffers a transition from a color string configuration between two
static color sources into a pair of static mesons due to the breaking of the
color string and the most favored decay into hadrons.

Among the different methods to solve the Schr\"odinger equation in order to 
find the quark-antiquark bound states, we use the Gaussian Expansion
Method~\cite{Hiyama2003} because it provides enough accuracy and it makes the
subsequent evaluation of the decay amplitude matrix elements easier. 

This procedure provides the radial wave function solution of the Schr\"odinger
equation as an expansion in terms of basis functions
\begin{equation}
R_{\alpha}(r)=\sum_{n=1}^{n_{max}} c_{n}^\alpha \phi^G_{nl}(r),
\end{equation} 
where $\alpha$ refers to the channel quantum numbers. The coefficients,
$c_{n}^\alpha$, and the eigenvalue, $E$, are determined from the Rayleigh-Ritz
variational principle
\begin{equation}
\sum_{n=1}^{n_{max}} \left[\left(T_{n'n}^\alpha-EN_{n'n}^\alpha\right)
c_{n}^\alpha+\sum_{\alpha'}
\ V_{n'n}^{\alpha\alpha'}c_{n}^{\alpha'}=0\right],
\end{equation}
where $T_{n'n}^\alpha$, $N_{n'n}^\alpha$ and $V_{n'n}^{\alpha\alpha'}$ are the 
matrix elements of the kinetic energy, the normalization and the potential, 
respectively. $T_{n'n}^\alpha$ and $N_{n'n}^\alpha$ are diagonal whereas the
mixing between different channels is given by $V_{n'n}^{\alpha\alpha'}$.

\begin{table}[!t]
\begin{center}
\begin{tabular}{ccc}
 \hline
 \hline
 \tstrutb
 Quark masses    & $m_{n}$ (MeV) & $313$ \\
 		 & $m_{s}$ (MeV) & $555$ \\
 		 & $m_{c}$ (MeV) & $1763$ \\
 		 & $m_{b}$ (MeV) & $5110$ \\[2ex]

 Goldstone Bosons & $m_{\pi}$ $(\mbox{fm}^{-1})$ & $0.70$ \\
		  & $m_{\sigma}$ $(\mbox{fm}^{-1})$ & $3.42$ \\
 		  & $m_{K}$ $(\mbox{fm}^{-1})$ & $2.51$ \\
 		  & $m_{\eta}$ $(\mbox{fm}^{-1})$ & $2.77$ \\
 		  & $\Lambda_{\pi}$ $(\mbox{fm}^{-1})$ & $4.20$ \\
		  & $\Lambda_{\sigma}$ $(\mbox{fm}^{-1})$ & $4.20$ \\
 		  & $\Lambda_{K}$ $(\mbox{fm}^{-1})$ & $4.21$ \\
		  & $\Lambda_{\eta}$ $(\mbox{fm}^{-1})$ & $5.20$ \\
 		  & $g^{2}_{ch}/4\pi$ & $0.54$ \\
 		  & $\theta_{p}$ $(^\circ)$ & $-15$ \\[2ex]

 OGE & $\alpha_{0}$ & $2.118$ \\
     & $\Lambda_{0}$ $(\mbox{fm}^{-1})$ & $0.113$ \\
     & $\mu_{0}$ (MeV) & $36.976$ \\
     & $\hat{r}_{0}$ (fm) & $0.181$ \\
     & $\hat{r}_{g}$ (fm) & $0.259$ \\[2ex]

 Confinement & $a_{c}$ (MeV) & $507.4$ \\
	       & $\mu_{c}$ $(\mbox{fm}^{-1})$ & $0.576$ \\
	       & $\Delta$ (MeV) & $184.432$ \\
	       & $a_{s}$ & $0.81$ \\
 \hline
 \hline
\end{tabular}
\caption{\label{tab:parameters} Quark model parameters.}
\end{center}
\end{table}

Following Ref.~\cite{Hiyama2003}, we employ Gaussian trial functions with
ranges  in geometric progression. This enables the optimization of ranges
employing a small number of free parameters. Moreover, the geometric
progression is dense at short distances, so that it allows the description of
the dynamics mediated by short range potentials. The fast damping of the
gaussian tail is not a problem, since we can choose the maximal range much
longer than the hadronic size.

Table~\ref{tab:parameters} shows the model parameters fitted over all meson
spectra and taken from Refs.~\cite{Vijande2005,Segovia2008}.

\section{WEAK DECAYS}
\label{sec:weakdecays}

In this section, we give an account of the semileptonic decays of the $B$ ($B$
or $B_s$) meson into orbitally excited charmed mesons. In the nonstrange sector,
this has been studied before within heavy quark effective theory (HQET) in
Refs.~\cite{Leibovich:1997tu,Leibovich:1997em}. There, only relative branching
ratios could be predicted and their results depended on the approximation used
and on two unknown functions, $\tau_1,\,\tau_2$, that describe corrections of
order $\Lambda_{QCD}/m_Q$. Only the ratio
$\Gamma^{\lambda=0}_{D^{**}}/\Gamma_{D^{**}}$, semileptonic decay rate with a
helicity 0 $D^{**}$ final meson over total semileptonic decay rate to that
meson, seemed to be  stable in the different approximations. We shall comment on
this below.

In the context of nonrelativistic constituent quark models, the state of a meson
is given by
\begin{align}
\left|\right.\!M,\lambda\vec P\!\left.\right>_{NR} = & \int 
\frac{d^3p}{(2\pi)^{3/2}} \sum_{\alpha_1,\alpha_2} 
\frac{(-1)^{1/2-s_{1}}}{\sqrt{2E_{f_1}(\vec p_1)2E_{f_2}(\vec p_2)}} \nonumber\\ 
&
\times  \hat{\phi}_{\alpha_1,\alpha_2}^{(M,\lambda)}(\vec{p}\,)
 \left|\right.\bar q,\alpha_1\,\vec p_1\left.\right>
\left|\right.q,\alpha_2\,\vec p_2 \left.\right>,
\end{align}
where $\vec P$ is the three-momentum of the meson and $\lambda$ is the spin
projection in the meson center of mass. The vector $\vec p$ is the relative
momentum of the $q\bar{q}$ pair, $ \vec p_1 =\frac{m_{f_1}}{m_{f_1}+m_{f_2}}\vec
P -\vec p $ and $\vec p_2 =\frac{m_{f_2}}{m_{f_1}+m_{f_2}}\vec P +\vec p$ are
the momenta of the antiquark and the quark, respectively, $\alpha_1$ and
$\alpha_2$ are the spin, flavor and color quantum numbers. $(E(\vec p_i),\vec
p_i)$ are the four-momenta and $m_i$ are the quark masses. The factor 
$(-1)^{1/2-s_1}$ is included in order that the antiquark spin states have the 
correct relative phase.

The normalization of the quark-antiquark states is
\begin{equation}
\left<\alpha'\,\vec p\,' | \alpha\, \vec p\,\right> = \delta_{\alpha',
  \alpha}\,(2\pi)^{3}\,2E_f(\vec p\,)\,\delta(\vec p\,' -\vec p),
\end{equation}
and the momentum space wave function $\hat
\phi_{\alpha_1, \alpha_2}^{(M,\lambda)}(\vec{p}\,)$ normalization is given by
\begin{equation}
\int d^3p\ \sum_{\alpha_1, \alpha_2}  (\hat
\phi_{\alpha_1,\alpha_2}^{(M,\lambda')}(\vec{p}\,))^{\ast}
\hat \phi_{\alpha_1,\alpha_2}^{(M,\lambda)}(\vec{p}\,)=\delta_{\lambda',\lambda}.
\end{equation}
Finally, the normalization of our meson states is
\begin{equation}
\label{norma1}
_{NR}\!\left\langle\right. \!\! M,\lambda'\vec{P}\,' | M,\lambda\vec{P} \! 
\left.\right\rangle_{NR} =
\delta_{\lambda',\lambda}(2\pi)^3\delta(\vec{P}'-\vec{P}).
\end{equation}

In the decay we have a $\bar{b} \to \bar{c}$ transition at the quark level and
we need to evaluate the hadronic matrix elements of the weak
current
\begin{equation}
J^{bc}_\mu (0)=\bar{\psi}_b(0)\gamma_\mu(I-\gamma_5)\psi_c(0).
\label{eq:current1}
\end{equation}
The hadronic matrix elements can be parameterized in terms of form factors as
\begin{widetext}
\begin{equation}
\label{FFdecomposition}
\begin{split}
&
\left\langle\right.\!\!D(0^{+}),\lambda\vec{P}_{D}|J_{\mu}^{bc}(0)|B(0^{-}),\vec
{P}_{B}\!\!\left.\right\rangle=P_{\mu}F_{+}(q^{2})+q_{\mu}F_{-}(q^{2}), \\
&
\begin{split}
\left\langle\right.\!\!D(1^{+}),\lambda\vec{P}_{D}|J_{\mu}^{bc}(0)|B(0^{-}),\vec
{P}_{B}\!\!\left.\right\rangle=&\frac{-1}{m_{B}+m_{D}}\epsilon_{
\mu\nu\alpha\beta }
\epsilon_{(\lambda)}^{\nu\ast}(\vec{P}_{D})P^{\alpha}q^{\beta}A(q^{2}) \\ 
&
-i\left\lbrace(m_{B}-m_{D})\epsilon_{(\lambda)\mu}^{\ast}(\vec{P}_{D})
V_{0}(q^{2})-\frac{P\cdot\epsilon_{(\lambda)}^{\ast}(\vec{P}_{D})}{m_{B}+m_{D}}
\left[P_{\mu}V_{+}(q^{2})+q_{\mu}V_{-}(q^{2})\right]\right\rbrace,
\end{split} \\
&
\begin{split}
\left<\right.\!\!D(2^+),\lambda \vec P_D\left| J_\mu^{bc}(0) \right| B(0^-)\vec
P_B\!\!\left.\right> =& \epsilon_{\mu\nu\alpha\beta}\epsilon^{\nu\delta*}_{
(\lambda)} (\vec P_D)P_\delta P^\alpha q^\beta T_4(q^2) \\ 
&-i\left\{\epsilon^*_{(\lambda)\mu\delta}(\vec P_D)P^\delta T_1(q^2)+
P^\nu P^\delta\epsilon^*_{(\lambda)\nu\delta}(\vec P_D) \left[P_\mu
T_2(q^2)+q_\mu T_3(q^{2})\right]\right\}.
\end{split}
\end{split}
\end{equation}
\end{widetext}
In the  expressions above, $P=P_B+P_D$ and $q=P_B-P_D$, $P_{B}$ and $P_{D}$
being the meson four-momenta. $m_{B}$ and $m_{D}$ are the meson masses,
$\epsilon^{\mu\nu\alpha\beta}$ is the fully antisymmetric tensor, for which the
convention $\epsilon^{0123}=+1$ is taken, and $\epsilon_{(\lambda)\mu}(\vec{P})$
and $\epsilon_{(\lambda)\mu\nu}(\vec{P})$ are the polarization vector and tensor
of vector and tensor mesons, respectively. The meson states in the Lorentz
decompositions of \req{FFdecomposition} are normalized such that
\begin{equation}
\left\langle\right. \!\! M,\lambda'\vec{P}\,' | M,\lambda\vec{P} \! 
\left.\right\rangle = \delta_{\lambda',\lambda}(2\pi)^3 2E_{M}(\vec{P})
\delta(\vec{P}'-\vec{P}).
\end{equation}
where $E_{M}(\vec{P})$ is the energy of the $M$ meson with three-momentum 
$\vec{P}$. Note the factor $2E_{M}$  difference with respect to \req{norma1}.

The form factors will be evaluated in the center of mass of the $0^-$ meson,
taking $\vec q$ in the $\hat z$ direction, so that $\vec P_B=\vec 0$ and $\vec
P_D = -\vec q=-|\vec{q}|\vec{k}$, with $\vec{k}$ representing the unit vector in
the $\hat z$ direction. We have taken the phases of the states such that all
form factors are real. $F_{+}$, $F_{-}$, $A$, $V_{0}$, $V_{+}$, $V_{-}$ and
$T_{1}$ are dimensionless, whereas $T_{2}$, $T_{3}$ and $T_{4}$ have dimension
of $E^{-2}$. Defining vector $V^\mu_\lambda(|\vec{q}\,|)$
 and axial $A^\mu_\lambda(|\vec{q}\,|)$ matrix elements  such that
\begin{align}
V^\mu_{\lambda}(|\vec q\,|)& =\left<\right.\!\! M_F,\lambda -|\vec q\,|\vec
k|J_V^{bc\mu}(0)|M_I,\vec 0 \!\left.\right>, \nonumber \\
A^\mu_{\lambda}(|\vec q\,|)& =\left<\right.\!\! M_F,\lambda -|\vec q\,|\vec
k|J_A^{bc\mu}(0)|M_I,\vec 0 \!\left.\right>,
\end{align} 
we have for a  $0^{-} \to 0^{+}$ decay, that the form factors are given in terms
of vector and axial matrix elements as
\begin{widetext}
\begin{equation}
\begin{split}
F_{+}(q^{2}) &= \frac{-1}{2m_{B}}
\left[A^{0}(|\vec{q}\,|)+\frac{A^{3}(|\vec{q}\,|)}{|\vec{q}\,|}
(E_{D}(-\vec{q}\,)-m_{B} )\right], \\
F_{-}(q^{2}) &= \frac{-1}{2m_{B}}
\left[A^{0}(|\vec{q}\,|)+\frac{A^{3}(|\vec{q}\,|)}{|\vec{q}\,|}
(E_{D}(-\vec{q}\,)+m_{B} )\right].
\end{split}
\end{equation}
\end{widetext}
In the case of a $0^- \to 1^+$ transition, the corresponding expressions for the
form factors are
\begin{widetext}
\begin{align}
&
A(q^{2})=-\frac{i}{\sqrt{2}}
\frac{m_{B}+m_{D}}{m_{B}|\vec{q}\,|}A_{\lambda=-1}^{1}(|\vec{q}\,|), \nonumber\\
&
V_{+}(q^{2})=+i\frac{m_{B}+m_{D}}{2m_{B}}
\frac{m_{D}}{|\vec{q}\,|m_{B}}\left\lbrace V_{\lambda=0}^{0}(|\vec{q}\,|)
-\frac{m_{B}-E_{D}(-\vec{q}\,)}{|\vec{q}\,|}V_{\lambda=0}^{3}(|\vec{q}\,|)
+\sqrt{2}\frac{m_{B}E_{D}(-\vec{q}\,)
-m_{D}^{2}}{|\vec{q}\,|m_{D}}V_{\lambda=-1}^{1}(|\vec{q}\,|)\right\rbrace, \nonumber \\
&
V_{-}(q^{2})=-i\frac{m_{B}+m_{D}}{2m_{B}}\frac{m_{D}}{|\vec{q}\,|m_{B}}
\left\lbrace -V_{\lambda=0}^{0}(|\vec{q}\,|)-\frac{m_{B}+E_{D}(-\vec{q}\,)}{|\vec{q}\,|}
V_{\lambda=0}^{3}(|\vec{q}\,|)
+\sqrt{2}\frac{m_{B}E_{D}(-\vec{q}\,)+m_{D}^{2}}{|\vec{q}\,|m_{D}}
V_{\lambda=-1}^{1}(|\vec{q}\,|)\right\rbrace, \nonumber\\
&
V_{0}(q^{2})=+i\sqrt{2}\frac{1}{m_{B}-m_{D}}V_{\lambda=-1}^{1}(|\vec{q}\,|).
\end{align}
\end{widetext}
Finally, the form factors for a  $0^- \to 2^+$ transition are given by the
relations
\begin{widetext}
\begin{align}
T_1(q^2)&= -i \frac{2m_D}{m_B|\vec q\,|}A_{T\lambda=+1}^1(|\vec q\,|), \nonumber\\ 
T_2(q^2)&= i \frac{1}{2m_B^3}\Bigg\{
-\sqrt{\frac32}\frac{m_D^2}{|\vec q\,|^2}A^0_{T\lambda=0}(|\vec q\,|)
-\sqrt{\frac32}\frac{m_D^2}{|\vec q\,|^3}(E_D(-\vec
q\,)-m_B)A^3_{T\lambda=0}(|\vec q\,|) \nonumber \\
&
 +\frac{2m_D}{|\vec
  q\,|} \left(1-\frac{E_D(-\vec q\,)(E_D(-\vec q\,)-m_B)}{|\vec
  q\,|^2}\right)A^1_{T\lambda=+1}(|\vec q\,|)
\Bigg\}, \nonumber \\ 
T_3(q^2)&= i\frac{1}{2m_B^3}\Bigg\{
-\sqrt{\frac32}\frac{m_D^2}{|\vec q\,|^2}A^0_{T\lambda=0}(|\vec q\,|)
-\sqrt{\frac32}\frac{m_D^2}{|\vec q\,|^3}(E_D(-\vec
q\,)+m_B)A^3_{T\lambda=0}(|\vec q\,|) \nonumber \\ 
&
 +\frac{2m_D}{|\vec
  q\,|}\left(1- \frac{E_D(-\vec q\,)(E_D(-\vec q\,)+m_B)}{|\vec
  q\,|^2}\right)A^1_{T\lambda=+1}(|\vec q\,|)
\Bigg\}, \nonumber\\ 
T_4(q^2)&= i\frac{m_D}{m_B^2|\vec q\,|^2}V_{T\lambda=+1}^1(|\vec q\,|).
\end{align}
\end{widetext}
The CQM evaluation of the vector and axial matrix elements 
$V_{\lambda}^{\mu}(|\vec{q}\,|)$ and $A_{\lambda}^{\mu}(|\vec{q}\,|)$ can be
found in the Appendix.

For a $B$ meson at rest and neglecting the neutrino mass, we have the  double 
differential decay width 
\begin{align}
\frac{d^2\Gamma}{dq^2dx_l}=&\frac{G^2_F}{64m_B^2}\frac{|V_{bc}|^2}{8\pi^3}
\frac{\lambda^{1/2}(q^2,m^2_B,m^2_D)}{2m_B}\frac{q^2-m_l^{2}}{q^2}\nonumber\\
&\times{\cal H}_{\alpha\beta}(P_B,P_D){\cal L}^{\alpha\beta}(p_l,p_\nu),
\end{align}
where $x_l$ is the cosine of the angle between the final meson momentum and the 
momentum of the final charged lepton measured in the lepton-neutrino center of
mass frame. $G_F=1.16637(1)\times10^{-5}\,\mbox{GeV}^{-2}$ is the Fermi
constant~\cite{PDG2010}, $m_{l}$ is the charged lepton mass,
$\lambda(a,b,c)=(a+b-c)^2-4ab$ and $V_{bc}$ is the $bc$ element of the
Cabbibo\,-\,Kobayashi\,-\,Maskawa matrix for which we shall use $V_{bc}=0.0413$.
${\cal H}_{\alpha\beta}$ and 
${\cal L}^{\alpha\beta}$ represent the hadron and lepton tensors. $P_B$, $P_D$,
$p_l$ and $p_\nu$ are the meson and lepton momenta.

Working  in the helicity formalism of Ref.~\cite{ivanov} and  after integration
on $x_l$ we have
\begin{align}
\frac{d\Gamma}{dq^{2}} =&
\frac{G_{F}^{2}}{8\pi^{3}}|V_{bc}|^{2}\frac{(q^{2}-m_{l}^{2})^{2}}
{12m_{B}^{2}q^{2}} \frac{\lambda^{1/2}(q^{2},m_{B}^{2},m_{D}^{2})}{2m_{B}} 
\nonumber \\ & \times(H_{U}+H_{L}+\tilde{H}_{U}+\tilde{H}_{L}+\tilde{H}_{S}),
\end{align}
where the suffixes ${U},{L},{S}$ stand for the unpolarized-transverse,
longitudinal and scalar components of the hadronic tensor, and
$\tilde{H}=\frac{m_l^2}{2q^2}H$. Integrating over $q^{2}$ we obtain the total
decay width that can be written as
\begin{equation}
\label{eq:semigam}
\Gamma = \Gamma_{U} + \Gamma_{L} + \tilde{\Gamma}_{U} + \tilde{\Gamma}_{L} +
\tilde{\Gamma}_{S},
\end{equation}
with $\Gamma_{J}$ and $\tilde{\Gamma}_{J}$ partial helicity widths defined as
\begin{align}
\Gamma_{J}=\int dq^{2}\frac{G_{F}^{2}}{8\pi^{3}}|V_{bc}|^{2}
\frac{(q^{2}-m_{l}^{2})^{2}}{12m_{B}^{2}q^{2}}
\frac{\lambda^{1/2}(q^{2},m_{B}^{2},m_{D}^{2})}{2m_{B}} H_{J}
\end{align}
and similarly for $\tilde{\Gamma}_{J}$ in terms of $\tilde{H}_{J}$.
The evaluation of the different form factors, and thus of the different helicity
amplitudes of the hadronic tensor, has been done following
Ref.~\cite{hnvv06}. 

\section{STRONG DECAYS}
\label{sec:strongdecays}

Meson strong decay is a complex nonperturbative process that has not yet been
described from QCD first principles. Instead, several phenomenological models
have been developed to deal with this topic, the $^{3}P_{0}$~\cite{mic69_1}, the
flux-tube~\cite{kok87_1}, and the Cornell~\cite{Eichten78,Swanson96} models
being the most popular.

Some models describe the decay process assuming that the extra quark-antiquark
pair is created from the vacuum. This is the case of the $^{3}P_{0}$ model,
which borrows its name from the quantum numbers of the created pair, or the
flux-tube model, which in addition to the creation vertex incorporates the
overlaps between the color flux tubes of the initial and final states. 

To address a more fundamental description of the decay mechanism, one has to 
describe hadron strong decays in terms of quark and gluon degrees of freedom. 
However, there has been little previous work in this area. Two different
examples are the study of open-charm decays of $c\bar{c}$ resonances by Eichten
{\it et al.}~\cite{Eichten78}, who assumed that the decays are due to pair
production from the static part of a Lorentz vector confining interaction,
and the study of a few strong decays in the light sector by Ackleh {\it et
al.}~\cite{Swanson96}, where the $q\bar{q}$ pair production comes
from the one-gluon exchange and a scalar confining interaction.

As we mentioned in the introduction, we shall use both the $^{3}P_{0}$ model and
a microscopic one, resembling those of Refs.~\cite{Eichten78}
and~\cite{Swanson96}, that originates from the different interaction pieces
present in  our interquark potential. These two approaches to meson production
are introduced in the following subsections.  

\subsection{The $^{3}P_{0}$ model}

It was first proposed by Micu~\cite{mic69_1} and further developed by Le
Yaouanc {\it et al.}~\cite{yao73_1}. To describe the meson decay process
$A\rightarrow B+C$, the $^{3}P_{0}$ model assumes that a quark-antiquark pair
is created with vacuum $J^{PC}=0^{++}$ quantum numbers. The created $q\bar q$ 
pair together with the $q\bar q$ pair present in the original meson regroup in
the two outgoing mesons via a quark rearrangement process.

The interaction Hamiltonian which describes the production process is given
by~\cite{Swanson96}
\begin{equation}
H_{I}=g\int d^{3}x \bar{\psi}(\vec{x})\psi(\vec{x})
\end{equation}
where $g$ is related to the dimensionless constant giving the strength of the 
$q\bar q$ pair creation from the vacuum as $\gamma=\frac{g}{2m_{q}}$, $m_{q}$
being the mass of the created quark. Note that the operator $g\bar{\psi}\psi$
leads to the decay $(q\bar{q})_{A}\to(q\bar{q})_{B}+(q\bar{q})_{C}$ through the
$a^{\dagger}b^{\dagger}$ term.

\subsection{The microscopic model}

In microscopic decay models one attempts to describe hadron strong decays in
terms of quark and gluon degrees of freedom. The quark-gluon decay mechanism
should give similar predictions to the reasonably accurate $^{3}P_{0}$ model and
should determine the strength of the $q\bar{q}$ pair creation, $\gamma$, of the
$^{3}P_{0}$ model in terms of more fundamental parameters.

Following Ref.~\cite{Swanson96}, the strong decays should be driven by the same
interquark Hamiltonian which determines the spectrum, the one-gluon exchange,
and the confining interaction appearing as the kernels. These interactions and
their associated decay amplitudes are undoubtedly all present and should be
added coherently. We already mentioned that our constituent quark model for the
heavy quark sector has a one-gluon exchange term and a mixture of Lorentz scalar
and vector confining interactions. This completely defines our microscopic model
for strong decays. Unlike previous works we use a screening confinement
interaction and also a mixture between scalar and vector Lorentz structures,
which is already fixed. 

The Hamiltonian of the interaction can be written as
\begin{equation}
\label{Hint}
H_{I}=\frac{1}{2}\int 
d^{3}\!xd^{3}\!y\,J^{a}(\vec{x})K(|\vec{x}-\vec{y}|)J^{a}(\vec{y}).
\end{equation}
The current $J^{a}$ in Eq.~(\ref{Hint}) is assumed to be a color octet. The
currents, $J$, with the color dependence $\lambda^{a}/2$ factored out and the
kernels, $K(r)$, for the interactions are
\begin{widetext}
\begin{itemize}
\item Currents
\begin{equation}
\label{currents}
J(\vec{x})=\bar{\psi}(\vec{x})\,\Gamma\,\psi(\vec{x})= \begin{cases} 
\bar{\psi}(\vec{x})\,\mathcal{I}\,\psi(\vec{x}) & \mbox{Scalar Lorentz current,}
\\
\bar{\psi}(\vec{x})\,\gamma^{0}\,\psi(\vec{x}) & \mbox{Static part of vector
Lorentz current,} \\
\bar{\psi}(\vec{x})\,\vec{\gamma}\,\psi(\vec{x}) & \mbox{Spatial part of 
vector Lorentz current,} \end{cases}
\end{equation}
\item Kernels
\begin{equation}
\label{kernels}
K(r)=\begin{cases} -4a_s\left[-a_{c}(1-e^{-\mu_{c}r})+\Delta\right] 
& \mbox{Confining interaction,} \\ 
+\frac{\alpha_{s}}{r} & \mbox{Color Coulomb OGE,} \\
-\frac{\alpha_{s}}{r} & \mbox{Transverse OGE.} \end{cases}
\end{equation}
\end{itemize}
\end{widetext}

For the Lorentz vector structure of the confinement we use 
$K(r)=\pm(1-a_s)4\left[-a_{c}(1-e^{-\mu_{c}r})+\Delta\right]$, where $\pm$ refers to 
static and transverse terms, respectively. We refer, following Ref.~\cite{Swanson96}, 
to this general type of 
interaction as a $JKJ$ decay model, and to the specific cases considered here 
as $sKs$, $j^{0}Kj^{0}$ and $j^{T}Kj^{T}$ interactions.

The wave functions for the mesons involved in the reactions are 
the solutions of the Schr\"odinger equation using the Gaussian Expansion 
Method mentioned above.  Details of the resulting matrix elements for  
different cases are given in Ref.~\cite{SegoviaF}.

\subsection{Strong decay width}

The total width is the sum over the partial widths characterized by the quantum 
numbers $J_{BC}$ and $l$
\begin{equation}
\Gamma_{A\rightarrow BC}=\sum_{J_{BC},l}\Gamma_{A\rightarrow BC}(J_{BC},l)
\end{equation}
where
\begin{equation}
\Gamma_{A\rightarrow BC}(J_{BC},l)=2\pi\int 
dk_{0}\delta(E_{A}-E_{BC})|\mathcal{M}_{A\rightarrow BC}(k_{0})|^{2}
\end{equation}
and $\mathcal{M}_{A\rightarrow BC}(k_{0})$ is calculated according to 
Refs.~\cite{Bonnaz1999,SegoviaF}.

Using relativistic phase space, we arrive at
\begin{equation}
\begin{split}
\Gamma_{A\rightarrow 
BC}(J_{BC},l)=2\pi\frac{E_{B}E_{C}}{m_
{A}k_{0}}|\mathcal{M}_{A\rightarrow BC}(k_{0})|^{2},
\end{split}
\end{equation}
where
\begin{equation}
k_{0}=\frac{\lambda^{1/2}(m_{A}^{2},m_{B}^2,m_{C}^{2})}{2m_{A}} 
\end{equation}
is the on shell relative momentum of mesons $B$ and $C$.

\section{RESULTS}
\label{sec:results}

For the low-lying positive parity excitations, any quark model predicts four
states that in the $^{2S+1}L_{J}$ basis correspond to $^1P_1$, $^3P_0$, $^3P_1$
and $^3P_2$. As charge conjugation is not well defined in the heavy-light
sector, $^1P_1$ and $^3P_1$ states can mix under the interaction.

In the infinite heavy quark mass limit, heavy quark symmetry (HQS) predicts two
degenerated $P$-wave meson doublets, labeled by $j_q=1/2$ with $J^P=0^+,1^+$
$(|1/2,0^+\rangle,|1/2,1^+\rangle)$ and $j_q=3/2$ with $J^P=1^+,2^+$
$(|3/2,1^+\rangle,|3/2,2^+\rangle)$. In this limit, the meson properties are
governed by the dynamics of the light quark, which is characterized by its total
angular momentum $j_q=s_q+L$, where $s_q$ is the light quark spin and $L$ the
orbital angular momentum. The total angular momentum of the meson $J$ is
obtained coupling $j_q$ to the heavy quark spin, $s_Q$. 

Moreover, in the infinite heavy quark mass limit the strong decays of the
$D_{J}\,(j_q=3/2)$  proceed only through $D$-waves, while the $D_{J}\,(j_q=1/2)$
decays happen only through $S$-waves~\cite{iw91}. The $D$-wave decay is
suppressed by the barrier factor which behaves as $q^{2L+1}$ where $q$ is the
relative momentum of the two decaying mesons. Therefore, the states decaying
through $D$-waves are expected to be narrower than those decaying via $S$-waves.

A change of basis allows to express the above states in terms of the
$^{2S+1}L_{J}$ basis, by recoupling angular momenta, as 
\begin{equation}
\begin{split}
&
|1/2,0^{+}\!\!\left.\right\rangle=+|^{3}
P_{0}\!\!\left.\right\rangle \\
&
\begin{split}
|1/2,1^{+}\!\!\left.\right\rangle
&=+\sqrt{\frac{1}{3}}|^{1}P_{1}\!\!\left.\right\rangle
 +\sqrt{\frac{2}{3}}|^{3}P_{1}\!\!\left.\right\rangle
\end{split} \\
&
\begin{split}
|3/2,1^{+} \!\!\left.\right\rangle
&=-\sqrt{\frac{2}{3}}|^{1}P_{1}\!\!\left.\right\rangle
 +\sqrt{\frac{1}{3}}|^{3}P_{1}\!\!\left.\right\rangle
\end{split} \\
&
|3/2,2^{+} \!\!\left.\right\rangle
=+|^{3}P_{2}\!\!\left.\right\rangle
\end{split}
\label{eq:mixing}
\end{equation}
where in the $^{2S+1}L_J$ wave functions we couple heavy and light quark spins,
in this order, to total spin $S$.

In the actual calculation the ideal mixing in Eq.~(\ref{eq:mixing}) between
$^1P_1$ and $^3P_1$ states changes due to finite charm quark mass effects. Our
CQM model  predicts the mixed states shown in Table~\ref{tab:mixedstates}, which
are very similar to the HQS states. This is expected since the $c$\,-\,quark is
much heavier ($m_{c}=1763\,{\rm MeV}$) than the light ($m_{n}=313\,{\rm MeV}$)
or strange ($m_{s}=555\,{\rm MeV}$) quarks. Note that now we have mixing,
even if small, between the $^{3}P_{2}$ and $^{3}F_{2}$ partial waves in  $2^{+}$
mesons. This is due to the OGE tensor term. 

In Ref.~\cite{Segovia2009} we have studied the $J^{P}=1^{+}$ charmed-strange
mesons, finding that the $J^{P}=1^{+}$ $D_{s1}(2460)$ has an important
non-$q\bar q$ contribution whereas the $J^{P}=1^{+}$ $D_{s1}(2536)$ is almost a
pure $q\bar{q}$ state. The presence of non-$q\bar q$ degrees of freedom in the 
$J^{P}=1^{+}$ charmed-strange meson sector enhances the $j_q=3/2$ component of
the $D_{s1}(2536)$. This wave function explains most of the experimental data,
as shown in Ref.~\cite{Segovia2009}, and it is the one we shall use here. For
this sector only the $q\bar q$ probabilities are given in
Table~\ref{tab:mixedstates}.

\begin{table}[t!]
\begin{center}
\begin{tabular}{ccccc}
\hline
\hline
\tstrutb
& $D_{0}^{\ast}$ & $D_{1}$ & $D'_{1}$ & $D_{2}^{\ast}$ \\[2ex]
$^{3}P_{0}$ & $+,\,1.0000$ & - & - & - \\
$^{1}P_{1}$ & - & $-,\,0.5903$ & $-,\,0.4097$ & - \\
$^{3}P_{1}$ & - & $+,\,0.4097$ & $-,\,0.5903$ & - \\
$^{3}P_{2}$ & - & - & - & $+,\,0.99993$ \\[2ex]
$1/2,0^+$ & $+,\,1.0000$ & - & - & - \\
$1/2,1^+$ & - & $+,\,0.0063$ & $-,\,0.9937$ & - \\
$3/2,1^+$ & - & $+,\,0.9937$ & $+,\,0.0063$ & - \\
$3/2,2^+$ & - & - & - & $+,\,0.99993$ \\
\hline
\tstrutb
& $D_{s0}^{\ast}$ & $D_{s1}$ & $D'_{s1}$ & $D_{s2}^{\ast}$ \\[2ex]
$^{3}P_{0}$ & $+,\,1.0000$ & - & - & - \\
$^{1}P_{1}$ & - & $-,\,0.7210$ & $-,\,0.1880$ & - \\
$^{3}P_{1}$ & - & $+,\,0.2770$ & $-,\,0.5570$ & - \\
$^{3}P_{2}$ & - & - & - & $+,\,0.99991$ \\[2ex]
$1/2,0^+$ & $+,\,1.0000$ & - & - & - \\
$1/2,1^+$ & - & $-,\,0.0038$ & $-,\,0.7390$ & - \\
$3/2,1^+$ & - & $+,\,0.9942$ & $-,\,0.0060$ & - \\
$3/2,2^+$ & - & - & - & $+,\,0.99991$ \\
\hline
\hline
\end{tabular}
\caption{\label{tab:mixedstates} Probability distributions  and their
relative phases for the four states predicted by CQM in the two basis
described in the text. In the $1^+$ strange sector the effects of
non-$q\bar q$ components are included, see text for details.}
\end{center}
\end{table}

\subsection{$B$ semileptonic decays into $D^{\ast\ast}$ mesons}

\subsubsection{Semileptonic $B \to D_{0}^{\ast}(2400)l\nu_{l}$ decay }

The measured branching fractions are ${\cal B}(B^+ \to \bar{D}^{\ast0}_{0} l^+
\nu_l){\cal B}(\bar{D}^{\ast0}_{0}\to D^{-}\pi^+)$ and ${\cal B}(B^0 \to
D^{\ast-}_{0} l^+ \nu_l){\cal B}(D^{\ast-}_{0}\to \bar{D}^{0}\pi^-)$. The meson
$D_{0}^{\ast}(2400)$ has $J^{P}=0^{+}$ quantum numbers and, therefore, due to
parity conservation, it decays only into $D\pi$, so that we have  ${\cal
B}(\bar{D}^{\ast0}_{0}\to D^{-}\pi^+)={\cal B}(D^{\ast-}_{0} \to
\bar{D}^{0}\pi^-)=2/3$ coming from isospin symmetry.

Table~\ref{tab:weakdecays0} shows the different helicity contributions to the
semileptonic width. In both cases the dominant contribution is given by
$\Gamma_{L}$ while the rest are negligible. The difference between the
semileptonic width of the charged and neutral $B$ meson is due to the large mass
difference between the $D_{0}^{\ast}$ and $D_{0}^{\ast\pm}$ mesons for which we
take the masses reported in Ref.~\cite{PDG2010}. 

\begin{table}[t!]
\begin{center}
\begin{tabular}{c|c|c}
\hline
\hline
\tstrutb
& $B^{+} \to \bar{D}_{0}^{\ast0}l^{+}\nu_{l}$ & $B^{0} \to
D_{0}^{\ast-}l^{+}\nu_{l}$ \\
\hline
\tstrutb
$\Gamma_{U}$ & $0.00$ & $0.00$ \\
$\tilde{\Gamma}_{U}$ & $0.00$ & $0.00$ \\
$\Gamma_{L}$ & $1.30$ & $1.16$ \\
$\tilde{\Gamma}_{L}$ & $6.83\times10^{-7}$ & $6.45\times10^{-7}$ \\ 
$\tilde{\Gamma}_{S}$ & $2.05\times10^{-6}$ & $1.93\times10^{-6}$ \\[2ex]
$\Gamma$ & $1.30$ & $1.16$ \\
\hline
\hline
\end{tabular}
\caption{\label{tab:weakdecays0} Helicity contributions and total decay widths, 
in units of $10^{-15}\,\mbox{GeV}$, for the $D_{0}^{\ast}$ meson.}
\end{center}
\end{table}

\begin{figure*}[t!]
\begin{center}
\parbox[c]{0.49\textwidth}{
\centering
\includegraphics[width=0.37\textwidth]{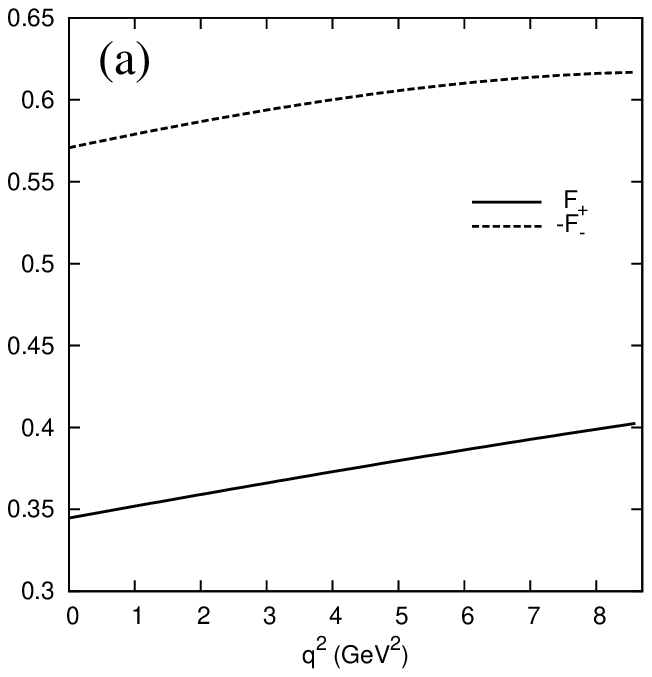}
}
\parbox[c]{0.49\textwidth}{
\centering
\includegraphics[width=0.40\textwidth]{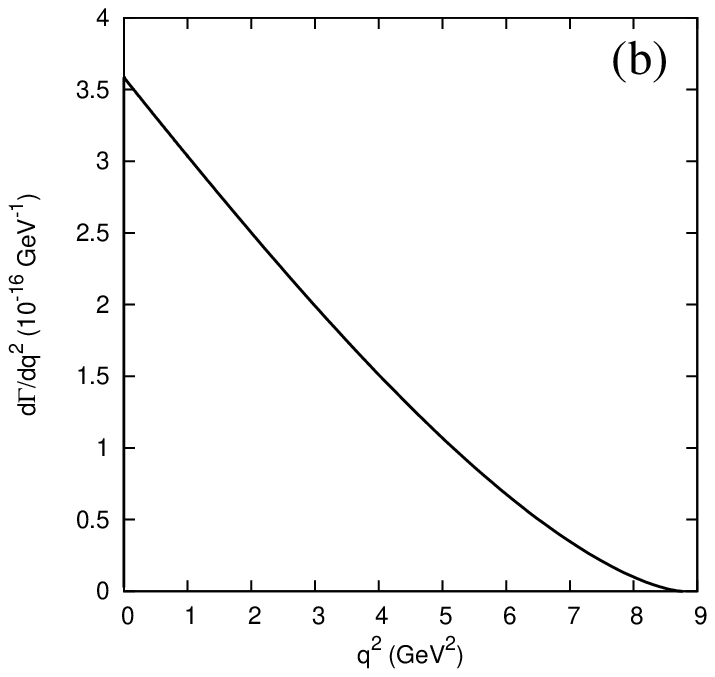}
}
\vspace*{8pt}
\caption{\label{fig:resD0z} Form factors and differential decay widths for the
$B^{+} \to \bar{D}_{0}^{\ast0}l^{+}\nu_{l}$ decay as a function of $q^{2}$. Very
similar results are obtained for the $B^{0} \to D_{0}^{\ast-}l^{+}\nu_{l}$
decay. {\bf (a)}: Form factors predicted by CQM. {\bf (b)}: Differential decay
width predicted by CQM.}
\end{center}
\end{figure*}

Figure~\ref{fig:resD0z} shows the $q^{2}$ dependence in the form factors and in
the differential decay width for ${\cal B}(B^+ \to \bar{D}^{\ast0}_{0} l^+
\nu_l)$, panels (a) and (b), respectively. Similar results (not shown) are
obtained for the ${\cal B}(B^0 \to D^{\ast-}_{0} l^+ \nu_l)$ case.

The final results for  the product of branching fractions are
\begin{equation}
\begin{split}
{\cal B}(B^+ \to \bar{D}^{\ast0}_{0} l^+ \nu_l){\cal B}(\bar{D}^{\ast0}_{0}\to
D^{-}\pi^+) &= 2.15 \times 10^{-3}, \\
{\cal B}(B^0 \to D^{\ast-}_{0} l^+ \nu_l){\cal B}(D^{\ast-}_{0}\to
\bar{D}^{0}\pi^-) &= 1.80 \times 10^{-3}, \\
\end{split}
\end{equation}
which compare very well with Belle data~\cite{belle}.

\subsubsection{Semileptonic $B \to D'_{1}(2430)l\nu_{l}$ decay }

The only Okubo\,-\,Zweig\,-\,Iizuka (OZI)-allowed decay channel for the $D'_{1}$
meson is $D'_{1}\to D^{\ast}\pi$ so that isospin symmetry predicts a branching
fraction $\mathcal{B}(D'_{1}\to D^{\ast}\pi^{\pm})=2/3$.

Table~\ref{tab:weakdecays1p} shows the different helicity contributions to the
semileptonic width of $B^{+} \to \bar{D}_{1}^{'0}l^{+}\nu_{l}$ and $B^{0} \to
D_{1}^{'-}l^{+}\nu_{l}$ calculated in the framework of the CQM. In this case,
$\Gamma_{U}$ and $\Gamma_{L}$ are of the same order of magnitude and give the
total semileptonic decay rate.

\begin{table}[t!]
\begin{center}
\begin{tabular}{c|c|c}
\hline
\hline
\tstrutb
& $B^{+} \to \bar{D}_{1}^{'0}l^{+}\nu_{l}$ & $B^{0} \to D_{1}^{'-}l^{+}\nu_{l}$
\\
\hline
\tstrutb
$\Gamma_{U}$ & $0.23$ & $0.23$ \\
$\tilde{\Gamma}_{U}$ & $1.35\times10^{-8}$ & $1.35\times10^{-8}$ \\
$\Gamma_{L}$ & $0.56$ & $0.56$ \\
$\tilde{\Gamma}_{L}$ & $4.12\times10^{-7}$ & $4.12\times10^{-7}$ \\ 
$\tilde{\Gamma}_{S}$ & $1.27\times10^{-6}$ & $1.27\times10^{-6}$ \\[2ex]
$\Gamma$ & $0.79$ & $0.80$ \\
\hline
\hline
\end{tabular}
\caption{\label{tab:weakdecays1p} Helicity contributions and total decay
widths, in units of $10^{-15}\,\mbox{GeV}$, for the $D'_{1}$ meson.}
\end{center}
\end{table}

\begin{figure*}[t!]
\begin{center}
\parbox[c]{0.49\textwidth}{
\centering
\includegraphics[width=0.37\textwidth]{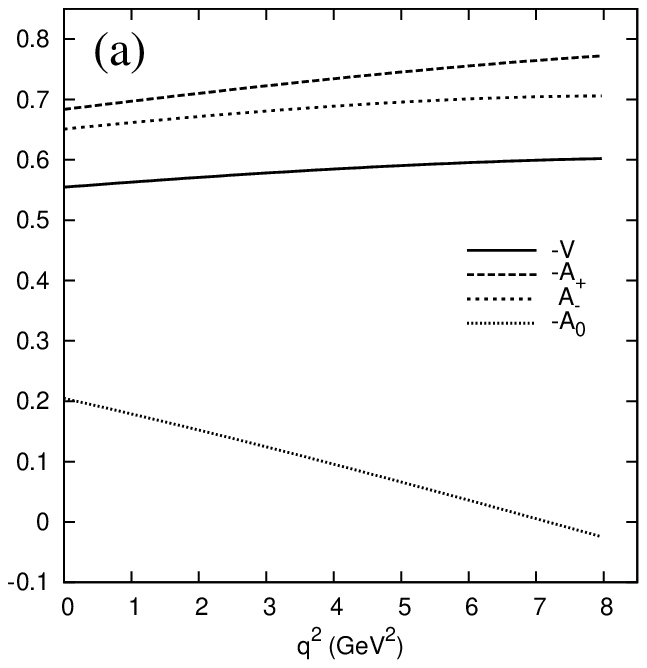}
}
\parbox[c]{0.49\textwidth}{
\centering
\includegraphics[width=0.40\textwidth]{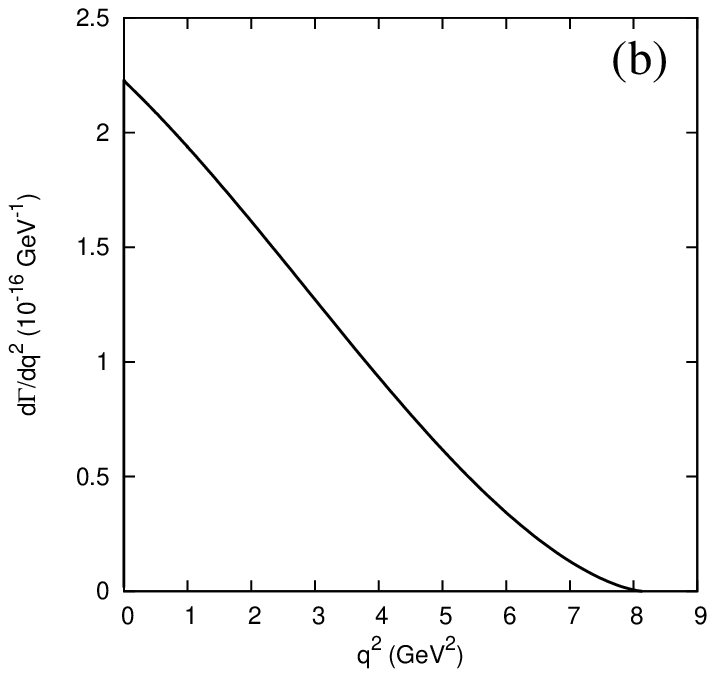}
}
\vspace*{8pt}
\caption{\label{fig:resD1pz} Form factors and differential decay widths for the
$B^{+} \to \bar{D}_{1}^{'0}l^{+}\nu_{l}$ decay as a function of $q^{2}$. Very
similar results are obtained for the $B^{0} \to D_{1}^{'-}l^{+}\nu_{l}$
decay. {\bf (a)}: Form factors predicted by CQM. {\bf (b)}: Differential decay
width predicted by CQM.}
\end{center}
\end{figure*}

Panels $(a)$ and $(b)$ of Fig.~\ref{fig:resD1pz} show the $q^{2}$ dependence of
the form factors and the differential decay width for the neutral 
$D'_{1}$ channel. A very similar result is obtained for the $D_{0}^{\ast}$ case.

We have in this case the  product of branching fractions
\begin{equation}
\begin{split}
{\cal B}(B^+ \to \bar{D}^{'0}_{1} l^+ \nu_l){\cal B}(\bar{D}^{'0}_{1}\to
D^{\ast-}\pi^+) &= 1.32 \times 10^{-3}, \\
{\cal B}(B^0 \to D^{'-}_{1} l^+ \nu_l){\cal B}(D^{'-}_1\to
\bar{D}^{\ast0}\pi^-) &= 1.23 \times 10^{-3}.
\end{split}
\end{equation}
which are a rough factor of $2$ smaller than the results from the BaBar
Collaboration~\cite{aubert08}.

\subsubsection{Semileptonic $B \to D_{1}(2420)l\nu_{l}$ decay }

As in the previous case, the branching fraction $\mathcal{B}(D_{1}\to
D^{\ast}\pi^{\pm})$ is again $2/3$ in our model because $D_{1}\to D^{\ast}\pi$
is the only OZI-allowed decay channel.

\begin{table}[t!]
\begin{center}
\begin{tabular}{c|c|c}
\hline
\hline
\tstrutb
& $B^{+}\rightarrow D_{1}^{0}l^{+}\nu_{l}$ & $B^{0}\rightarrow
D_{1}^{-}l^{+}\nu_{l}$ \\
\hline
\tstrutb
$\Gamma_{U}$ & $0.38$ & $0.38$ \\
$\tilde{\Gamma}_{U}$ & $1.94\times10^{-8}$ & $1.93\times10^{-8}$ \\
$\Gamma_{L}$ & $1.17$ & $1.16$ \\
$\tilde{\Gamma}_{L}$ & $7.16\times10^{-7}$ & $7.15\times10^{-7}$ \\ 
$\tilde{\Gamma}_{S}$ & $2.17\times10^{-6}$ & $2.17\times10^{-6}$ \\[2ex]
$\Gamma$ & $1.55$ & $1.54$ \\
\hline
\hline
\end{tabular}
\caption{\label{tab:weakdecays1} Helicity contributions and total decay widths, 
in units of $10^{-15}\,\mbox{GeV}$, for the $D_{1}$ meson.}
\end{center}
\end{table}

Table~\ref{tab:weakdecays1} shows the different helicity contributions to the
semileptonic width of the reactions $B^{+} \to \bar{D}_{1}^{0}l^{+}\nu_{l}$ and
$B^{0} \to \bar{D}_{1}^{-}l^{+}\nu_{l}$. The most
important contribution is given by $\Gamma_{L}$. The ratio
$\Gamma_{L}/\Gamma=0.75$ gives the probability for the final $D_1$ meson to have
helicity 0. This result is in agreement with the values $0.72-0.81$ obtained in
the HQET calculation of Ref.~\cite{Leibovich:1997em}. 

\begin{figure*}[t!]
\begin{center}
\parbox[c]{0.49\textwidth}{
\centering
\includegraphics[width=0.37\textwidth]{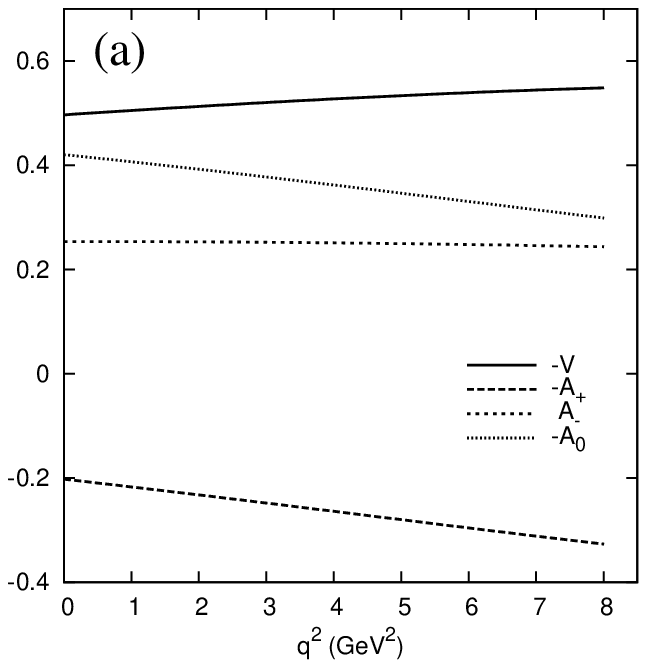}
}
\parbox[c]{0.49\textwidth}{
\centering
\includegraphics[width=0.40\textwidth]{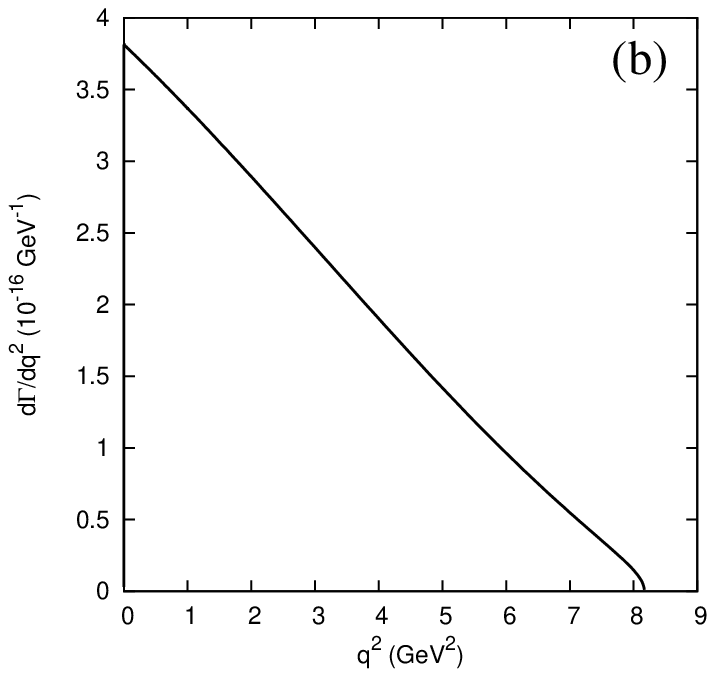}
}
\vspace*{8pt}
\caption{\label{fig:resD1z} Form factors and differential decay widths for the
$B^{+} \to D_{1}^{0}l^{+}\nu_{l}$ decay as a function of $q^{2}$. The
differences with respect $B^{0} \to D_{1}^{-}l^{+}\nu_{l}$ are negligible. {\bf
(a)}: Form factors predicted by CQM. {\bf (b)}: Differential decay width
predicted by CQM.}
\end{center}
\end{figure*}

Figure~\ref{fig:resD1z} shows the $q^{2}$ dependence of the form factors and
the differential decay width for neutral $D_{1}$ channel, in panels $(a)$ and
$(b)$, respectively. Again, a very similar result is obtained for  the charged
case.

The product of branching fractions are 
\begin{align}
{\cal B}(B^+ \to \bar{D}^0_1 l^+ \nu_l){\cal B}(\bar{D}^0_1 \to
D^{\ast-}\pi^+)=2.57\times10^{-3}, \nonumber \\
{\cal B}(B^0 \to D^-_1 l^+ \nu_l){\cal B}(D^-_1\to
\bar{D}^{\ast0}\pi^-)=2.39\times10^{-3},
\end{align}
which in this case compare very well with the latest BaBar data~\cite{aubert09}.

\subsubsection{Semileptonic $B \to D_{2}^{\ast}l\nu_{l}$ decay }

The semileptonic decay is studied by reconstructing the decay channel 
$D_{2}^{\ast} \to D^{(\ast)}\pi^{-}$, using the decay chain $D^{\ast}\to
D^{0}\pi$ for $D^{\ast}$ meson and $D^{0}\to K^{-}\pi^{+}$ or $D^{+} \to
K^{-}\pi^{+}\pi^{+}$ for $D$ meson. What is actually measured is the product of
branching fractions ${\cal B}(B^+ \to \bar{D}^{\ast0}_2 l^+\nu_l){\cal
B}(\bar{D}^{\ast0}_2\to D^{-}\pi^+)$ and ${\cal B}(B^+ \to \bar{D}^{\ast0}_2 l^+
\nu_l){\cal B}(\bar{D}^{\ast0}_2\to D^{\ast-}\pi^+)$.

The first step of this decay involves a semileptonic process which can be 
calculated using \req{eq:semigam}. In Table~\ref{tab:weakdecays2} we show the 
different helicity contributions to the total width. The main contribution is 
$\Gamma_{L}$ in both neutral and charged $D_{2}^{\ast}$ channels, providing 
almost $2/3$ of the total width. The following one is $\Gamma_{U}$, the
rest of the contributions  being negligible. Again our ratio
$\Gamma_L/\Gamma=0.67$ is in agreement with the values $0.63-0.64$ obtained in
Ref.~\cite{Leibovich:1997em} using HQET.

\begin{table}[t!]
\begin{center}
\begin{tabular}{c|c|c}
\hline
\hline
\tstrutb
& $B^{+}\rightarrow D_{2}^{\ast0}l^{+}\nu_{l}$ & $B^{0}\rightarrow
D_{2}^{\ast-}l^{+}\nu_{l}$ \\
\hline
\tstrutb
$\Gamma_{U}$ & $0.44$ & $0.44$ \\
$\tilde{\Gamma}_{U}$ & $2.56\times10^{-8}$ & $2.56\times10^{-8}$ \\
$\Gamma_{L}$ & $0.90$ & $0.91$ \\
$\tilde{\Gamma}_{L}$ & $5.27\times10^{-7}$ & $5.29\times10^{-7}$ \\ 
$\tilde{\Gamma}_{S}$ & $1.54\times10^{-6}$ & $1.55\times10^{-6}$ \\[2ex]
$\Gamma$ & $1.34$ & $1.35$ \\
\hline
\hline
\end{tabular}
\caption{\label{tab:weakdecays2} Helicity contributions and total decay widths, 
in units of $10^{-15}\,\mbox{GeV}$, for the $D_{2}^{\ast}$ meson.}
\end{center}
\end{table}

\begin{figure*}[t!]
\begin{center}
\parbox[c]{0.49\textwidth}{
\centering
\includegraphics[width=0.37\textwidth]{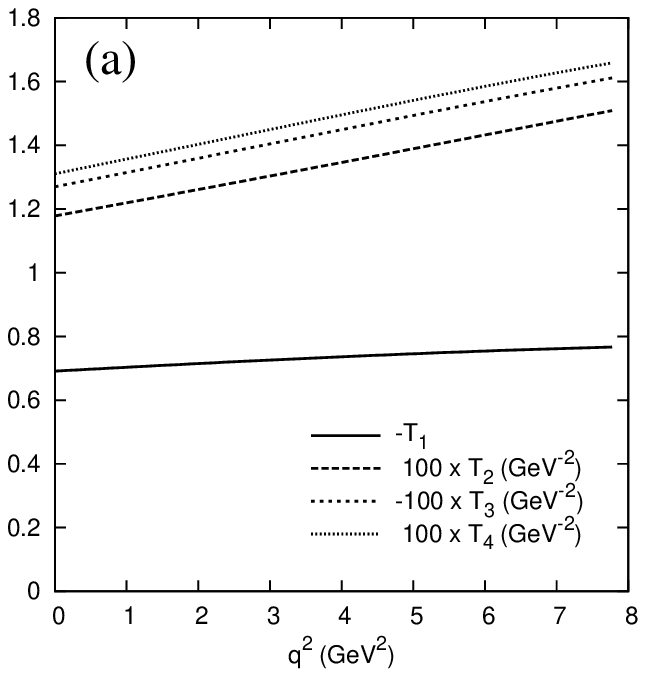}
}
\parbox[c]{0.49\textwidth}{
\centering
\includegraphics[width=0.40\textwidth]{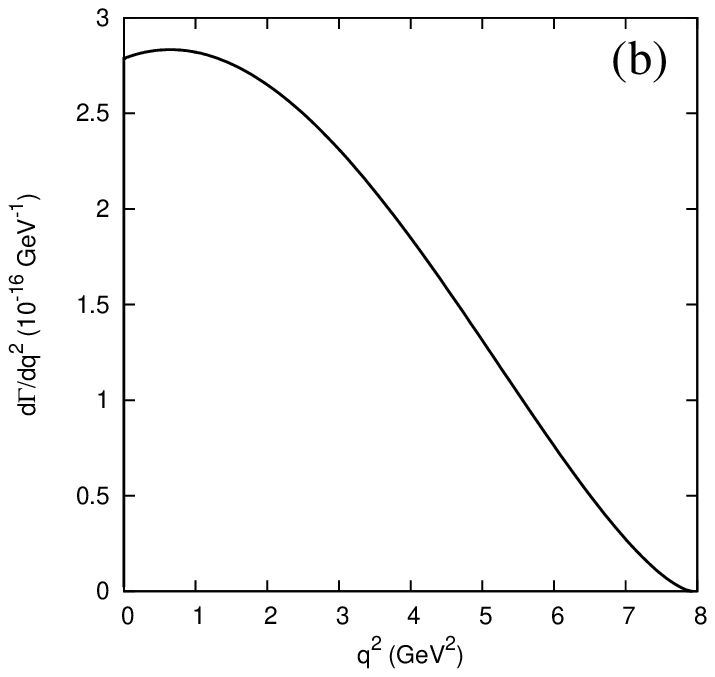}
}
\vspace*{8pt}
\caption{\label{fig:resD2z} Form factors and differential decay widths for the
$B^{+} \to D_{2}^{\ast0}l^{+}\nu_{l}$ decay as a function of $q^{2}$. Very
similar results are obtained for the $B^{0} \to D_{2}^{\ast-}l^{+}\nu_{l}$
decay. {\bf (a)}: Form factors predicted by CQM. {\bf (b)}: Differential decay
width predicted by CQM.}
\end{center}
\end{figure*}

Figure~\ref{fig:resD2z} shows 
the $q^{2}$ dependence in the form factors and in the differential decay width, 
panels (a) and (b) respectively, for the $B^{+}\to \bar{D}_{2}^{\ast0}l^{+}
\nu_{l}$ decay. Very similar results (not shown) are obtained for the $B^{0} \to
D_{2}^{\ast-}l^{+}\nu_{l}$ case.

The subsequent strong decays which appear are $D_{2}^{\ast} \to D^{\ast}\pi^{-}$ 
and $D_{2}^{\ast}\to D\pi^{-}$. In Table~\ref{tab:ratiosD2} we show the strong
decay branching ratios obtained with the $^3P_0$ and microscopic models. They
are in good agreement with experimental data~\cite{PDG2010}.

\begin{table}[t!]
\begin{center}
\begin{tabular}{lccc}
\hline
\hline
\tstruta
Branching ratio & Exp. & $^{3}P_{0}$ & Microscopic \\
\hline
\tstrutb
$\Gamma(D^{0}\pi^{+})/\Gamma(D^{\ast0}\pi^{+})$ & $1.9\pm1.1\pm0.3$ & $1.80$ &
$1.97$ \\
$\Gamma(D^{+}\pi^{-})/\Gamma(D^{\ast+}\pi^{-})$ & $1.56\pm0.16$ & $1.82$ &
$1.97$ \\
$\Gamma(D^{+}\pi^{-})/\Gamma(D^{(\ast)+}\pi^{-})$ & $0.62\pm0.03\pm0.02$ &
$0.65$ & $0.66$ \\
\hline
\hline
\end{tabular}
\caption{\label{tab:ratiosD2} Branching ratios for $D_{2}^{\ast}$ decays
collected by the PDG~\cite{PDG2010} and our theoretical results calculated
through the two strong decay models.}
\end{center}
\end{table}

Finally, we obtain the products of branching fractions for both decay chains
considering that the total width of the $D_{2}^{\ast}$ meson is the sum of the
partial widths of $D^{\ast}\pi$ and $D\pi$ channels since these are
the only OZI-allowed processes

\begin{align}
{\cal B}(B^+ \to D^{\ast0}_2 l^+ \nu_l){\cal B}(D^{\ast0}_2\to
D^{+}\pi^-)=&\left\lbrace\begin{matrix} 1.44\times10^{-3} \\
1.48\times10^{-3} \end{matrix}\right. \nonumber \\
{\cal B}(B^+ \to D^{\ast0}_2 l^+ \nu_l){\cal B}(D^{\ast0}_2\to
D^{\ast+}\pi^-)=&\left\lbrace\begin{matrix} 0.79\times10^{-3} \\
0.75\times10^{-3} \end{matrix}\right. \nonumber \\
{\cal B}(B^0 \to D^{\ast-}_2 l^+ \nu_l){\cal B}(D^{\ast-}_2\to
D^{0}\pi^-)=&\left\lbrace\begin{matrix} 1.34\times10^{-3} \\
1.38\times10^{-3} \end{matrix}\right. \nonumber \\
{\cal B}(B^0 \to D^{\ast-}_2 l^+ \nu_l){\cal B}(D^{\ast-}_2\to
D^{\ast0}\pi^-)=&\left\lbrace\begin{matrix} 0.74\times10^{-3} \\
0.70\times10^{-3} \end{matrix}\right. \nonumber \\
\end{align}
where the first one refers to the calculation using the $^{3}P_{0}$ model and
the second one comes from the microscopic model. These results are in very good
agreement with  BaBar data~\cite{aubert09}. 

\subsubsection{Summary of the results }

Final results and their comparisons with the experimental data are given in
Table~\ref{tab:experiment2}. Except for the $D_{1}'(2430)$, the predictions are
in very good agreement with the latest experimental measurements, Belle for
$D_{0}(2400)$ and BaBar for $D_{1}(2420)$ and $D_{2}^{\ast}(2460)$. For the
$D_{1}'(2430)$ there is also a strong disagreement between  experimental data
in the neutral case.

\begin{table*}[t!]
\begin{center}
\begin{tabular}{lcccc}
\hline
\hline
\tstrutb
& Belle~\cite{belle} & BaBar~\cite{aubert08,aubert09} & $^{3}P_{0}$ & Mic. \\
& $(\times10^{-3})$ & $(\times10^{-3})$ & $(\times10^{-3})$
& $(\times10^{-3})$ \\
\hline
\tstrutb
$D_{0}^{\ast}(2400)$ & & & & \\[2ex]
${\cal B}(B^+ \to \bar{D}^{\ast0}_{0} l^+ \nu_l){\cal B}(\bar{D}^{\ast0}_{0}\to
D^{-}\pi^+)$ & $2.4\pm0.4\pm0.6$ & $2.6\pm0.5\pm0.4$ & $2.15$ & $2.15$ \\
${\cal B}(B^0 \to D^{\ast-}_{0} l^+ \nu_l){\cal B}(D^{\ast-}_{0}\to
\bar{D}^{0}\pi^-)$ & $2.0\pm0.7\pm0.5$ & $4.4\pm0.8\pm0.6$ & $1.80$ & $1.80$
\\
\hline
\tstrutb
$D'_{1}(2430)$ & & & & \\[2ex]
${\cal B}(B^+ \to \bar{D}^{'0}_{1} l^+ \nu_l){\cal B}(\bar{D}^{'0}_{1}\to
D^{\ast-}\pi^+)$ & $<0.7$ & $2.7\pm0.4\pm0.5$ & $1.32$ & $1.32$ \\
${\cal B}(B^0 \to D^{'-}_{1} l^+ \nu_l){\cal B}(D^{'-}_1\to
\bar{D}^{\ast0}\pi^-)$ & $<5$ & $3.1\pm0.7\pm0.5$ & $1.23$ & $1.23$ \\
\hline
\tstrutb
$D_{1}(2420)$ & & & & \\[2ex]
${\cal B}(B^+ \to \bar{D}^0_1 l^+ \nu_l){\cal B}(\bar{D}^0_1\to D^{\ast-}\pi^+)$
& $4.2\pm0.7\pm0.7$ & $2.97\pm0.17\pm0.17$ & $2.57$ & $2.57$ \\
${\cal B}(B^0 \to D^-_1 l^+ \nu_l){\cal B}(D^-_1\to \bar{D}^{\ast0}\pi^-)$ &
$5.4\pm1.9\pm0.9$ & $2.78\pm0.24\pm0.25$ & $2.39$ & $2.39$
\\
\hline
\tstrutb
$D_{2}^{\ast}(2460)$ & & & & \\[2ex]
${\cal B}(B^+ \to \bar{D}^{\ast0}_2 l^+ \nu_l){\cal B}(\bar{D}^{\ast0}_2\to
D^{-}\pi^+)$ & $2.2\pm0.3\pm0.4$ & $1.4\pm0.2\pm0.2^{(\ast)}$ & $1.43$ &
$1.47$ \\
${\cal B}(B^+ \to \bar{D}^{\ast0}_2 l^+ \nu_l){\cal B}(\bar{D}^{\ast0}_2\to
D^{\ast-}\pi^+)$ & $1.8\pm0.6\pm0.3$ & $0.9\pm0.2\pm0.2^{(\ast)}$ & $0.79$ &
$0.75$ \\
${\cal B}(B^+ \to \bar{D}^{\ast0}_2 l^+ \nu_l){\cal B}(\bar{D}^{\ast0}_2\to
D^{(\ast)-}\pi^+)$ & $4.0\pm0.7\pm0.5$ & $2.3\pm0.2\pm0.2$ & $2.22$ &
$2.22$ \\[2ex]
${\cal B}(B^0 \to D^{\ast-}_2 l^+ \nu_l){\cal B}(D^{\ast-}_2\to
\bar{D}^{0}\pi^-)$ & $2.2\pm0.4\pm0.4$ & $1.1\pm0.2\pm0.1^{(\ast)}$ & $1.34$
& $1.38$ \\
${\cal B}(B^0 \to D^{\ast-}_2 l^+ \nu_l){\cal B}(D^{\ast-}_2\to
\bar{D}^{\ast0}\pi^-)$ & $<3$ & $0.7\pm0.2\pm0.1^{(\ast)}$ & $0.74$ & $0.70$
\\
${\cal B}(B^0 \to D^{\ast-}_2 l^+ \nu_l){\cal B}(D^{\ast-}_2\to
\bar{D}^{(\ast)0}\pi^-)$ & $<5.2$ & $1.8\pm0.3\pm0.1$ & $2.08$ &
$2.08$ \\[2ex]
${\cal B}_{D/D^{(\ast)}}$ & $0.55\pm0.03$ & $0.62\pm0.03\pm0.02$ & $0.65$ &
$0.66$ \\
\hline
\hline
\end{tabular}
\caption{\label{tab:experiment2} Most recent experimental measurements reported
by Belle and BaBar Collaborations and their comparison with our results. The
symbol $(\ast)$ indicates the estimated results from the original data using
$B_{D/D^{(\ast)}}$.}
\end{center}
\end{table*}

\subsection{$B_{s}$ semileptonic decays into $D_{s}^{\ast\ast}$ mesons}

The semileptonic decays of $B_{s}$ meson into orbitally excited $P$-wave
charmed-strange mesons ($D_{s}^{\ast\ast}$) provides an extra opportunity to get
more insight into this system.

The $j_{q}=1/2$ doublet, $D_{s0}^{\ast}(2318)$ and $D_{s1}(2460)$, shows
surprisingly light masses which are below the $DK$ and $D^{\ast}K$ thresholds,
respectively. These unexpected properties have triggered many theoretical
interpretations, including four quark states, molecules, and the coupling of the
$q\bar{q}$ components with different structures. As mentioned before, the 
$D_{s1}(2460)$ meson has an important non-$q\bar q$ contribution. 

\begin{figure}[t!]
\begin{center}
\parbox[c]{0.49\textwidth}{
\centering
\includegraphics[width=0.40\textwidth]{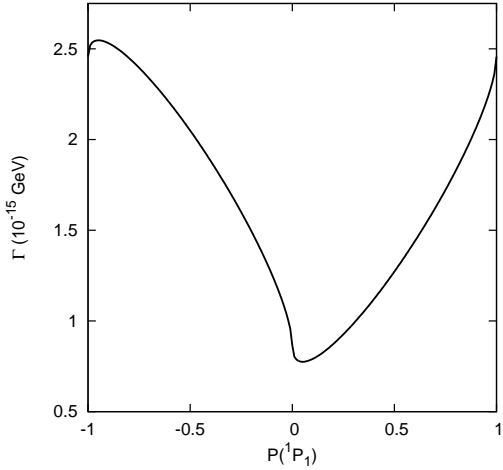}
}
\vspace*{8pt}
\caption{\label{fig:resDs12460} Decay width for the $B_{s}^{0} \to 
D_{s1}(2460)^{-}\mu^{+}\nu_{\mu}$ decay as a function of the $^{1}P_{1}$
component
probability. The sign reflects the relative phase between 
$^{1}P_{1}$ and $^{3}P_{1}$ components: -1 opposite phase and +1 same phase.}
\end{center}
\end{figure}

We have calculated the semileptonic $B_{s}$ decays assuming that the
$D_{s}^{\ast\ast}$ mesons are pure $q\bar{q}$ systems. For the
$D_{s0}^\ast(2318)$ and $D_{s1}(2460)$, which are below the corresponding
$D^{(*)}K$ thresholds, we only quote the weak decay branching fractions.
Concerning the $D_{s1}(2460)$, and as shown in Ref.~\cite{Segovia2009}, the
$^{1}P_{1}$ and $^{3}P_{1}$ probabilities change with the coupling to non-$q\bar
q$ degrees of freedom. What we do here is to vary these probabilities (including
the phase) in order to obtain the limits of the decay width in the case of the
$D_{s1}(2460)$ being a pure $q\bar{q}$ state, see Fig.~\ref{fig:resDs12460}.
Assuming that non-$q\bar{q}$ components will give a small contribution to the
weak decay, experimental results lower than these limits  will be an indication
of a more complex structure for this meson.

\begin{table*}[t!]
\begin{center}
\begin{tabular}{lccc}
\hline
\hline
\tstrutb
& Experiment & \multicolumn{2}{c}{Theory} \\
& $(\times10^{-3})$ & \multicolumn{2}{c}{$(\times10^{-3})$} \\
\hline
\tstrutb
$D_{s0}^{\ast}(2318)$ & & & \\[2ex]
${\cal B}(B_{s}^{0} \to D^{\ast}_{s0}(2318)^{-} \mu^+ \nu_\mu)$ & - &
\multicolumn{2}{c}{$4.43$} \\
\hline
\tstrutb
$D_{s1}(2460)$ & & & \\[2ex]
${\cal B}(B_{s}^{0} \to D_{s1}(2460)^{-} \mu^+ \nu_\mu)$ & - &
\multicolumn{2}{c}{$1.74-5.70$} \\
\hline
\tstrutb
$D_{s1}(2536)$ & & $^{3}P_{0}$ & Mic. \\[2ex]
${\cal B}(B_{s}^{0} \to D_{s1}(2536)^{-} \mu^+ \nu_\mu ){\cal
B}(D_{s1}(2536)^{-}\to D^{\ast-}\bar{K}^{0})$ &
$2.4\pm0.7$~\cite{PDG2010,abazov09} & $2.05$ &
$2.24$ \\
\hline
\tstrutb
$D_{s2}^{\ast}(2573)$ & & $^{3}P_{0}$ & Mic. \\[2ex]
${\cal B}(B_{s}^0 \to D^{\ast}_{s2}(2573)^{-} \mu^+ \nu_\mu){\cal
B}(D^{\ast}_{s2}(2573)^{-} \to D^{-}\bar{K}^{0})$ & - & $1.70$ & $1.77$ \\
${\cal B}(B_{s}^0 \to D^{\ast}_{s2}(2573)^{-} \mu^+ \nu_\mu){\cal
B}(D^{\ast}_{s2}(2573)^{-} \to D^{\ast-}\bar{K}^{0})$ & - & $0.18$ & $0.11$
\\
${\cal B}(B_{s}^0 \to D^{\ast}_{s2}(2573)^{-} \mu^+ \nu_\mu){\cal
B}(D^{\ast}_{s2}(2573)^{-} \to D^{(\ast)-}\bar{K}^{0})$ & - & $1.88$ &
$1.88$ \\
\hline
\hline
\end{tabular}
\caption{\label{tab:experiment3}  Our predictions and their comparison with
the available experimental data for semileptonic $B_{s}$ decays into orbitally
excited charmed-strange mesons.}
\end{center}
\end{table*}

For the decay into $D_{s1}(2536)$, our model predicts  the weak decay branching
fraction ${\cal B}(B_s^0\to D_{s1}(2536)\mu^+ \nu_\mu)=4.77\times 10^{-3}$ and
the strong branching fractions ${\cal B}(D_{s1}(2536)^{-}\to
D^{\ast-}\bar{K}^{0}) = 0.43\, (0.47)$ for the $^{3}P_{0}$ (microscopic) models.
The final result appears in Table~\ref{tab:experiment3}. It is in good agreement
with the  existing experimental data~\cite{PDG2010}, which to us is a
confirmation of our former result in Ref.~\cite{Segovia2009} about the
$q\bar{q}$ nature of this state.

In the case of the $D_{s2}^{\ast}(2573)$ the open strong decays are $DK$ and
$D^{\ast}K$, so the experimental measurements must be referred to ${\cal
B}(B_{s}^0 \to D^{\ast}_{s2}(2573)^{-} \mu^+ \nu_\mu){\cal
B}(D^{\ast}_{s2}(2573)^{-} \to D^{-}\bar{K}^{0})$ and ${\cal B}(B_{s}^0 \to
D^{\ast}_{s2}(2573)^{-} \mu^+ \nu_\mu){\cal B}(D^{\ast}_{s2}(2573)^{-} \to
D^{\ast-}\bar{K}^{0})$. 

For the weak branching fraction we get in this case ${\cal B}(B_{s}^{0} \to
D^{\ast}_{s2}(2573)^{-} \mu^+\nu_\mu)=3.76\times10^{-3}$. For the strong decay
part of the reaction, we obtain in our model
\begin{equation}
\begin{split}
{\cal B}(D_{s2}^{\ast-}\to D^{-}\bar{K}^{0}) &= \begin{cases} 0.45 \\ 0.47,
\end{cases} \\
{\cal B}(D_{s2}^{\ast-}\to D^{\ast-}\bar{K}^{0}) &= \begin{cases} 0.047 \\
0.030, \end{cases} \\
\end{split}
\end{equation}
where the first one refers to the calculation using the $^{3}P_{0}$ model and
the second one comes from the microscopic model. Our final results can be
seen in Table.~\ref{tab:experiment3}.

Besides we predict the ratio
\begin{equation}
\frac{\Gamma(D_{s2}^{\ast}\to DK)}{\Gamma(D_{s2}^{\ast}\to
DK)+\Gamma(D_{s2}^{\ast}\to D^{\ast}K)} = \begin{cases} 0.91 & ^3P_0 \\
0.94 & {\rm Mic.} \end{cases}
\end{equation}

\section{Conclusions}
\label{sec:conclusions}

We have performed a calculation of the branching fractions for the semileptonic
decays of $B$ and $B_{s}$ mesons into final states containing orbitally excited
charmed and charmed-strange mesons, respectively. 

We worked in the framework of the constituent quark model of
Ref.~\cite{Vijande2005}. The model parameters were fitted to the
meson spectra in Refs.~\cite{Vijande2005, Segovia2008}. Our meson states are
close to the ones predicted by HQS as expected.

We have calculated the semileptonic decay rates within the helicity formalism of
Ref.~\cite{ivanov} and following the work in Ref.~\cite{hnvv06}. The strong
decay widths have been calculated using two models, the $^{3}P_{0}$ model and a
microscopic model based on the quark-antiquark interactions present in the CQM
model of Ref.~\cite{Vijande2005}.

From the experimental point of view, Belle and BaBar Collaborations provide
their most recent measurements for the $B$ meson in Refs.~\cite{belle}
and~\cite{aubert08,aubert09} respectively. For the $B_{s}$ meson  only the
product of branching fractions ${\cal B}(B_{s}^{0} \to D_{s1}(2536)^{-} \mu^+
\nu_\mu){\cal B}(D_{s1}(2536)^{-} \to D^{\ast-}\bar{K}^{0})$ has been
determined~\cite{PDG2010} using the experimental data on  
${\cal B}(\bar b\to B_s^0){\cal B}(B_{s}^{0} \to D_{s1}(2536)^{-} \mu^+
\nu_\mu){\cal B}(D_{s1}(2536)^{-}\to D^{\ast-}\bar{K}^{0})$ measured by the D0 
Collaboration~\cite{abazov09} and the PDG's best value for ${\cal B}(\bar b\to
B_s^0)$~\cite{PDG2010}.

Our results for $B$ semileptonic decays into $D_{0}^{\ast}(2400)$, $D_{1}(2420)$
and $D_{2}(2460)$ are in good agreement with the latest experimental
measurements. In the case of $D'_{1}(2430)$ the prediction lies a factor of $2$
below BaBar data. Note however the disagreement between BaBar and Belle data for
the neutral case.

In the case of $B_{s}$ semileptonic decays, our prediction for the ${\cal
B}(B_{s}^{0} \to D_{s1}(2536)^{-} \mu^+ \nu_\mu){\cal B}(D_{s1}(2536)^{-}\to
D^{\ast-}\bar{K}^{0})$ product of branching fractions is in agreement with the
experimental data. This, together with the properties calculated in
Ref.~\cite{Segovia2009}, is to us evidence of a dominant $q\bar q$ structure for
the $D_{s1}(2536)$ meson. We also  give predictions for decays into other
$D_s^{**}$ mesons which can be useful to test the $q\bar q$ nature of these
states.

\begin{acknowledgments}

This work has been partially funded by the Spanish Ministerio de Ciencia y Tecnolog\'ia
under Contracts Nos. FIS2006-03438,  FIS2009-07238 and FPA2010-21750-C02-02, by 
the Spanish Ingenio-Consolider 2010
Programs CPAN CSD2007-00042 and MultiDark CSD2009-0064, and  by 
the European Community-Research
Infrastructure Integrating Activity 'Study of Strongly Interacting Matter'
(HadronPhysics2 Grant No. 227431). C. A. thanks a Juan de la Cierva contract from the
Spanish  Ministerio de Educaci\'on y Ciencia.

\end{acknowledgments}

\appendix

\begin{widetext}

\section{Form factor decomposition of hadronic matrix elements}

Here we give general expressions valid for transitions between a pseudoscalar
meson $M_{I}$ at rest with quark content $\bar {q}_{f_{1}}q_{f_{2}}$ and a final
$M_{F}$ meson with total angular momentum and parity $J^{P}=0^+,1^{+},2^+$, 
three-momentum $-|\vec{q}\,|\vec{k}$, and quark content
$\bar {q}_{f'_{1}}q_{f_{2}}$. The transition changes the antiquark 
flavor. Following Ref.~\cite{hnvv06}  we evaluate 
$V^\mu_{\lambda}(|\vec{q}\,|)$ and
$A^\mu_{\lambda}(|\vec{q}\,|)$ in the CQM through the relations

\begin{align}
V^\mu_{\lambda}(|\vec q\,|)& =\sqrt{2m_I2E_F(-\vec
q\,)}\,\,
_{NR}\!\left<\right. M_F,\lambda -|\vec q\,|\vec k|J_V^{bc\mu}(0)|M_I,\vec 0
\left. \right>_{NR} \nonumber \\
\nonumber\\
A^\mu_{\lambda}(|\vec q\,|)& =\sqrt{2m_I2E_F(-\vec
q\,)}\,\,
_{NR}\!\left<\right. M_F,\lambda -|\vec q\,|\vec k|J_A^{bc\mu}(0)|M_I,\vec 0
\left. \right>_{NR}
\end{align}
For the different cases under study we will have the following.

\subsection{Case $0^- \to 0^+$}

\begin{equation}
\begin{split}
A^{0}(|\vec{q}\,|)=& 
\sqrt{2m_{I}2E_{F}(-\vec{q}\,)} \int d^{3}p
\frac{1}{4\pi|\vec{p}\,|}\left(\hat{\phi}_{f'_{1}f_{2}}^{(M(0^{+}))}(|\vec{p}\,
|)\right)^{\ast}\hat{\phi}_{f_{1}f_{2}}^{(M(0^{-}))}\left(|\vec{p}-\frac{m_{f_{
2}}}{m_{f'_{1}}+m_{f_{2}} } q\vec{k}|\right) \\
&
\sqrt{\frac{\hat{E}_{f'_{1}}\hat{E}_{f_{1}}}{4E_{f'_{1}}E_{f_{1}}}}\left[\frac{
\vec{p}\cdot\left(\frac{m_{f_{2}}}{m_{f'_{1}}+m_{f_{2}}}|\vec{q}\,|\vec{k}
-\vec{p}\right)}{\hat{E}_{f_{1}}}+\frac{
\vec{p}\cdot\left(-\frac{m_{f'_{1}}}{m_{f'_{1}}+m_{f_{2}}}|\vec{q}\,|\vec{k}
-\vec{p}\right)}{\hat{E}_{f_{1}}}\right], \\
A^{3}(|\vec{q}\,|)=& 
\sqrt{2m_{I}2E_{F}(-\vec{q}\,)} \\
&
\times\int
d^{3}p\frac{1}{4\pi|\vec{p}\,|}\left(\hat{\phi}_{f'_{1}f_{2}}^{(M(0^{+}))}
(|\vec{p}\,|)\right)^{\ast}\hat{\phi}_{f_{1}f_{2}}^{(M(0^{-}))}\left(|\vec{p}
-\frac{m_{f_{
2}}}{m_{f'_{1}}+m_{f_{2}} }
q\vec{k}|\right)\sqrt{\frac{\hat{E}_{f'_{1}}\hat{E}_{f_{1}}}{4E_{f'_{1}}E_{f_{1}
}}} \\
&
\times\Bigg\lbrace
p_{z}\left(1-\frac{\left(-\frac{m_{f'_{1}}}{m_{f'_{1}}+m_{f_{2}}}
|\vec{q}\,|\vec{k}-\vec{p}\right)\cdot\left(\frac{m_{f_{2}}}{m_{f'_{1}}+m_{f_{2}
} } |\vec{q}\,|\vec{k}-\vec{p}\right)}{\hat{E}_{f'_{1}}\hat{E}_{f_{1}}}
\right)+\frac{1}{\hat{E}_{f'_{1}}\hat{E}_{f_{1}}}\left[\left(-\frac{m_{f'_{1}}}{
m_{f'_{1}}+m_{f_{2}}}|\vec{q}\,|-p_{z}\right) \right. \\
&
\left. \times \vec{p} \cdot
\left(\frac{m_{f_{2}}}{m_{f'_{1}}+m_{f_{2}}}|\vec{q}\,|\vec{k}-\vec{p}
\right)+\left(\frac{m_{f_{2}}}{m_{f'_{1}}+m_{f_{2}}}|\vec{q}\,|-p_{z}
\right)\vec{p}\cdot\left(-\frac{m_{f'_{1}}}{m_{f'_{1}}+m_{f_{2}}}
|\vec{q}\,|\vec{k}
-\vec{p}\right) \right] \Bigg\rbrace.
\end{split}
\end{equation}
$E_{f'_1}$ and $E_{f_1}$ are shorthand notations for
$E_{f'_1}(-\frac{m_{f'_1}}{m_{f'_1}+m_{f_2}}
|\vec{q}\,|\vec{k}-\vec{p}\,)$ and $E_{f_1}(\frac{m_{f_2}}{m_{f'_1}+m_{f_2}}
|\vec{q}\,|\vec{k}-\vec{p}\,)$ respectively and $\hat{E}_{f}=E_{f}+m_{f}$.

\subsection{Case $0^- \to 1^+$}

Here we have to distinguish two different cases that depend on the total spin
$S$ of the quark-antiquark system.

\begin{enumerate}[i)]

\item Case $S=0$
\begin{align}
V_{\lambda=0}^{(1^{+},S=0)0}(|\vec{q}\,|) =&
-i\sqrt{3}\sqrt{2m_{I}2E_{F}(-\vec{q}\,)}\int\,d^{3}p\sqrt{\frac{\hat{E}_{
f'_{1}}\hat{E}_{f_{1}}
}{4E_{f'_{1}}E_{f_{1}}}}\frac{1}{4\pi
p}\left(\hat{\phi}_{f'_{1}f_{2}}^{(M_{F}(1^{+},S=0))}(p)\right)^{\ast}
\nonumber \\ 
&\times\hat{\phi}_{f_{1}f_{2}}^{(M_{I}(0^{-}))}(|\vec{p}-\frac{m_{f_{2}}}{m_{f'_
{1}}+m_{f_{2}}}|\vec{q}\,|\vec{k}\,|)\,\,p_{z}\left[1+\frac{\left(-\frac{m_{f'_{
1}}}{m_{f'_{1
}}+m_{f_{2}}}|\vec{q}\,|\vec{k}-\vec{p}\,\right)\cdot\left(\frac{m_{f_{2}}}{m_{
f'_{1}}+m_{f_{2}}}|\vec{q}\,|\vec{k}-\vec{p}\,\right)}{\hat{E}_{f'_{1}}\hat{E}_{
f_{1}}}\right], \nonumber \\ 
V_{\lambda=-1}^{(1^{+},S=0)1}(|\vec{q}\,|) =&
i\sqrt{\frac{3}{2}}\sqrt{2m_{I}2E_{F}(-\vec{q}\,)}\int\,d^{3}p\sqrt{\frac
{\hat{E}_{f'_{1}}\hat{E}_{f_{1}}}
{4E_{f'_{1}}E_{f_{1}}}}\frac{1}{
4\pi p}\left(\hat{\phi}_{f'_{1}f_{2}}^{(M_{F}(1^{+},S=0))}(p)\right)^{\ast}
\nonumber \\
&
\times\hat{\phi}_{f_{1}f_{2}}^{(M_{I}(0^{-}))}(|\vec{p}-\frac{m_{f_{2}}}{m_{f'_{
1}}+m_{f_{2}}}|\vec{q}\,|\vec{k}\,|)\,\,p_{x}^{2}\left(\frac{1}{\hat{E}_{f_{1}}}
+\frac{1}{
\hat{E}_{f'_{1}}}\right), \nonumber \\
V_{\lambda=0}^{(1^{+},S=0)3}(|\vec{q}\,|) =&
-i\sqrt{3}\sqrt{2m_{I}2E_{F}(-\vec{q}\,)}\int\,d^{3}p\sqrt{\frac{\hat{E}_{
f'_{1}}\hat{E}_{f_{1}}}
{4E_{f'_{1}}E_{f_{1}}}}\frac{1}{4\pi
p}\left(\hat{\phi}_{f'_{1}f_{2}}^{(M_{F}(1^{+},S=0))}(p)\right)^{\ast}
\nonumber \\
&
\times\hat{\phi}_{f_{1}f_{2}}^{(M_{I}(0^{-}))}(|\vec{p}-\frac{m_{f_{2}}}{m_{f'_{
1}}+m_{f_{2}}}|\vec{q}\,|\vec{k}\,|)\,\,p_{z}\left(\frac{\frac{m_{f_{2}}}{m_{f'_
{1}}+m_{f_{2}
}}|\vec{q}\,|-p_{z}}{\hat{E}_{f_{1}}}-\frac{\frac{m_{f'_{1}}}{m_{f'_{1}}+m_{f_{2
}}}|\vec{q}\,|+p_{z}}{\hat{E}_{f'_{1}}}\right), \nonumber \\
A_{\lambda=-1}^{(1^{+},S=0)1}(|\vec{q}\,|) =&
-i\sqrt{\frac{3}{2}}\sqrt{2m_{I}2E_{F}(-\vec{q}\,)}\int\,d^{3}p\sqrt{\frac
{\hat{E}_{f'_{1}}\hat{E}_{f_{1}}}
{4E_{f'_{1}}E_{f_{1}}}}\frac{1}{
4\pi p}\left(\hat{\phi}_{f'_{1}f_{2}}^{(M_{F}(1^{+},S=0))}(p)\right)^{\ast}
\nonumber\\
&
\times\hat{\phi}_{f_{1}f_{2}}^{(M_{I}(0^{-}))}(|\vec{p}-\frac{m_{f_{2}}}{m_{f'_{
1}}+m_{f_{2}}}|\vec{q}\,|\vec{k}\,|)\frac{p_{y}^{2}|\vec{q}\,|}{\hat{E}_{f_{1}}
\hat{E}_{f'_{1}}}.\\\nonumber
\end{align}

\item Case $S=1$

\begin{align}
V_{\lambda=0}^{(1^{+},S=1)0}(|\vec{q}\,|) =&
i\sqrt{\frac{3}{2}}\sqrt{2m_{I}2E_{F}(-\vec{q}\,)}\int\,d^{3}p\sqrt{\frac{
\hat{E}_{f'_{1}}\hat{E}_{f_{1}}} {4E_{f'_{1}}E_{f_{1}}}}\frac{1}{4\pi
p}\left(\hat{\phi}_{f'_{1}f_{2}}^{(M_{F}(1^{+}, S=1))}(p)\right)^{\ast}
\nonumber \\
&
\times\hat{\phi}_{f_{1}f_{2}}^{(M_{I}(0^{-}))}(|\vec{p}-\frac{m_{f_{2}}}{m_{f'_{
1}}+m_{f_{2}}}|\vec{q}\,|\vec{k}\,|)\,\frac{|\vec{q}\,|(p_{z}^{2}-p^{2})}{\hat{E
}_{f'_{1}} \hat{E}_{f_{1}}}, \nonumber\\
V_{\lambda=-1}^{(1^{+},S=1)1}(|\vec{q}\,|)=&-i\frac{\sqrt{3}}{2}
\sqrt{2m_{I}2E_{F}(-\vec{q}\,)}\int\,d^{3}p\sqrt{\frac
{\hat{E}_{f'_{1}}\hat{E}_{f_{1}}}{4E_{f'_{1}}E_{f_{1}}}}\frac{1}{
4\pi p}\left(\hat{\phi}_{f'_{1}f_{2}}^{(M_{F}(1^{+},S=1))}(p)\right)^{\ast}
\nonumber\\
&\times\hat{\phi}_{f_{1}f_{2}}^{(M_{I}(0^{-}))}(|\vec{p}-\frac{m_{f_{2}}}{m_{f'_
{1}}+m_{f_{2}}}|\vec{q}\,|\vec{k}\,|)\left(\frac{p_{y}^{2}+p_{z}^{2}+p_{z}|\vec{
q}\,|\frac{m_
{f'_{1}}}{m_{f'_{1}}+m_{f_{2}}}}{\hat{E}_{f'_{1}}}-\frac{p_{y}^{2}+p_{z}^{2}-p_{
z}|\vec{q}\,|\frac{m_{f_{2}}}{m_{f'_{1}}+m_{f_{2}}}}{\hat{E}_{f_{1}}}\right),
\nonumber\\
V_{\lambda=0}^{(1^{+},S=1)3}(|\vec{q}\,|) =&
i\sqrt{\frac{3}{2}}\sqrt{2m_{I}2E_{F}(-\vec{q}\,)}\int\,d^{3}p\sqrt{\frac{
\hat{E}_{f'_{1}}\hat{E}_{f_{1}}}
{4E_{f'_{1}}E_{f_{1}}}}\frac{1}{4\pi
p}\left(\hat{\phi}_{f'_{1}f_{2}}^{(M_{F}(1^{+},S=1))}(p)\right)^{\ast}
\nonumber\\
&
\times\hat{\phi}_{f_{1}f_{2}}^{(M_{I}(0^{-}))}(|\vec{p}-\frac{m_{f_{2}}}{m_{f'_{
1}}+m_{f_{2}}}|\vec{q}\,|\vec{k}\,|)\,(p_{x}^{2}+p_{y}^{2})\left(\frac{1}{\hat{E
}_{f_{1}}}-\frac{1}{\hat{E}_{f'_{1}}}\right), \nonumber \\
A_{\lambda=-1}^{(1^{+},S=1)1}(|\vec{q}\,|) =&
i\frac{\sqrt{3}}{2}\sqrt{2m_{I}2E_{F}(-\vec{q}\,)} \nonumber\\
&\int\,d^{3}p\sqrt{\frac{\hat{E}_{f'_{1}}\hat{E}_{f_{1}}}{4E_{f'_{1}}E_{f_{1}}}}
\frac{1}{4\pi
p}\left(\hat{\phi}_{f'_{1}f_{2}}^{(M_{F}(1^{+},S=1))}(p)\right)^{\ast}
\hat{\phi}_{f_{1 }f_{2}}^{(M_{I}(0^{-}))}(|\vec{p}-\frac{m_{f_{2}}}{m_{
f'_{1}}+m_{f_{2}}}|\vec{q}\,|\vec{k}\,|) \nonumber\\
& 
\times
\left\lbrace
p_{z}\left[1-\frac{\left(-\frac{m_{f'_{1}}}{m_{f'_{1}}+m_{f_{2}}}|\vec{q}\,|\vec
{k}-\vec{p}\,\right)\cdot\left(\frac{m_{f_{2}}}{m_{f'_{1}}+m_{f_{2}}}|\vec{q}\,
|\vec{k}-\vec{p}\,\right)}{\hat{E}_{f'_{1}}\hat{E}_{f_{1}}}\right]+\frac{m_{f_{2
}}-m_{f'_{1}}}{m_{f'_{1}}+m_{f_{2}}}\frac{p_{x}^{2}|\vec{q}\,|}{\hat{E}_{f'_{1}}
\hat{E}_{f_{1}}}\right\rbrace. \\\nonumber
\end{align}

\end{enumerate}

\newpage

\subsection{Case $0^- \to 2^+$}

Here we have to distinguish between $L=1$ and $L=3$.

\begin{enumerate}[i)]

\item Case $L=1$

\begin{align}
V^{(2^+,L=1)1}_{\lambda=+1}(|\vec
q\,|)=&-i\frac{\sqrt3}{2}
\sqrt{2m_I2E_F(-\vec q\,)}\int d^3p\sqrt{\frac{\hat
E_{f'_1}\hat E_{f_1}}{4E_{f'_1}E_{f_1}}}\frac{1}{4\pi p}
\left(\hat \phi ^{(M_F(2^+,L=1))}_{f'_1 f_2}(p)\right)^* \nonumber \\
&
\times \hat{\phi}^{(M_I(0^-))}_{f_1 f_2}(|\vec p - \frac{m_{f_2}}{m_{f'_1}
+m_{f_2}}|\vec q\,|\vec k|) \left(\frac{p_y^2-p_z^2-p_z|\vec
q\,|\frac{m_{f'_1}}{m_{f'_1}+m_{f_2}}}{\hat E_{f'_1}}-
\frac{p_y^2-p_z^2+p_z|\vec q\,|\frac{m_{f_2}}{m_{f'_1}+m_{f_2}}}{\hat E_{f_1}}
\right), \nonumber \\
A^{(2^+,L=1)0}_{\lambda=0}(|\vec q\,|)=&-\frac{i}{\sqrt2} \sqrt{2m_I2E_F(-\vec
q\,)}\int d^3p\sqrt{\frac{\hat E_{f'_1}\hat
E_{f_1}}{4E_{f'_1}E_{f_1}}}\frac{1}{4\pi p} \left(\hat \phi
^{(M_F(2^+,L=1))}_{f'_1 f_2}(p)\right)^* \nonumber \\
&
\times \hat{\phi}^{(M_I(0^-))}_{f_1 f_2}(|\vec p - \frac{m_{f_2}}{m_{f'_1}
+m_{f_2}}|\vec q\,|\vec k|)
\left(\frac{p_x^2+p_y^2-2p_z^2-2p_z|\vec q\,|\frac{m_{f'_1}}{m_{f'_1}+m_{f_2}}}
{\hat E_{f'_1}} \right. \nonumber \\
&\left.+\frac{p_x^2+p_y^2-2p_z^2+2p_z|\vec
q\,|\frac{m_{f_2}}{m_{f'_1}+m_{f_2}}}{\hat
E_{f_1}}\right), \nonumber \\
A^{(2^+,L=1)1}_{\lambda=+1}(|\vec q\,|)=&
i\frac{\sqrt3}{2}\sqrt{2m_I2E_F(-\vec q\,)}\int d^3p \sqrt{\frac{\hat
E_{f'_1}\hat E_{f_1}}{4E_{f'_1}E_{f_1}}}\frac{1}{4\pi p}
\left(\hat \phi ^{(M_F(2^+,L=1))}_{f'_1 f_2}(p)\right)^* \nonumber \\
&
\times \hat{\phi}^{(M_I(0^-))}_{f_1 f_2}(|\vec p - \frac{m_{f_2}}{m_{f'_1}
+m_{f_2}}|\vec q\,|\vec k\,|)
\left\lbrace p_z\left[1-\frac{\left(-\frac{m_{f'_1}}{m_{f'_1} +m_{f_2}}|\vec
q\,|\vec k -\vec p\,\right)\cdot
\left(\frac{m_{f_2}}{m_{f'_1} +m_{f_2}}|\vec q\,|\vec k-\vec p\,\right)}
{\hat E_{f'_1}\hat E_{f_1}}\right] \right. \nonumber \\
& 
\left. +\frac{4p_zp_x^2-p_x^2|\vec q\,|
\frac{m_{f_2}-m_{f'_1}}{m_{f'_1} +m_{f_2}}}{\hat E_{f'_1}\hat E_{f_1}}
\right\rbrace, \nonumber \\
A^{(2^+,L=1)3}_{\lambda=0}(|\vec q\,|)=&-i{\sqrt2}\sqrt{2m_I2E_F(-\vec q\,)}\int
d^3p\sqrt{\frac{\hat E_{f'_1}\hat E_{f_1}}{4E_{f'_1}E_{f_1}}}\frac{1}{4\pi p}
\left(\hat \phi ^{(M_F(2^+,L=1))}_{f'_1 f_2}(p)\right)^* \nonumber \\
&
\times \hat{\phi}^{(M_I(0^-))}_{f_1 f_2}(|\vec p - \frac{m_{f_2}}{m_{f'_1}
+m_{f_2}}|\vec q\,|\vec k\,|) \Bigg\lbrace
p_z\Bigg[1-\frac{\left(-\frac{m_{f'_1}}{m_{f'_1} +m_{f_2}}|\vec q\,|\vec k -\vec
p\,\right)\cdot
\left(\frac{m_{f_2}}{m_{f'_1} +m_{f_2}}|\vec q\,|\vec k-\vec p\,\right)}
{\hat E_{f'_1}\hat E_{f_1}}\Bigg] \nonumber \\
&
+\frac{1}{\hat E_{f'_1}\hat E_{f_1}} \left[ 2p_z\left(-\frac{m_{f'_1}}{m_{f'_1}
+m_{f_2}}|\vec q\,| -p_z \right) \left(\frac{m_{f_2}}{m_{f'_1} +m_{f_2}}|\vec
q\,|-p_z \right)+ \right.\nonumber\\ 
&
\left. (p_x^2+p_y^2)\left(-p_z+\frac{m_{f_2}-m_{f'_1}}{2(m_{f'_1}
+m_{f_2})}|\vec q\,|\right) \right]\Bigg\rbrace. \\ \nonumber
\end{align}

\item Case $L=3$

\begin{align}
V_{\lambda=+1}^{(2^{+},L=3)1}(|\vec{q}\,|)=&
\frac{i}{\sqrt{8}}\sqrt{2m_{I}2E_{F}(-\vec{q}\,)}\int
d^{3}p\sqrt{\frac{\hat{E}_{f'_{1}}\hat{E}_{f_{1}}}{4E_{f'_{1}}E_{f_{1}}}}\frac{1
}{4\pi p^{3}}
\left(\hat{\phi}_{f'_{1}f_{2}}^{(M(2^{+},L=3))}(
p)\right)^{\ast} \nonumber \\
&
\times\hat{\phi}_{f_{1}f_{2}}^{(M(0^{-}))}
(|\vec{p}-\frac{m_{f_{2}}}{m_{f'_{1}}+m_{f_{2}}}|\vec q\,|\vec{k}\,|)
\nonumber \\
&
\times\Bigg[\frac{1}{\hat{E}_{f_{1}}}\Bigg(p^{2}\Big(2p^{2}_{y}-3p_{z}\Big(\frac
{m_{ f_{2 } } } { m_
{f'_{1}}+m_{f_{2}}}|\vec{q}\,|-p_{z}\Big)\Big) \nonumber \\ &
+5p_{z}\Big(-2p^{2}_{y}p_{z}+\Big(\frac{m_{f_{2}}}{m_{f'_{1}}+m_{f_{2}}}|\vec{q}
\, |-p_ { z }\Big)(p^{2}_{x}-p^{2}_{y}+p^{2}_{z})\Big)\Bigg) \nonumber \\
&
+\frac{1}{\hat{E}_{f'_{1}}}\Bigg(p^{2}\Big(-2p^{2}_{y}+3p_{z}\Big(-\frac{m_{f'_{
1}}} {m_{
f'_{1}}+m_{f_{2}}}|\vec{q}\,|-p_{z}\Big)\Big) \nonumber \\ &
-5p_{z}\Big(-2p^{2}_{y}p_{z}+\Big(-\frac{m_{f'_{1}}}{m_{f'_{1}}+m_{f_{2}}}|\vec{
q}\, |-p_
{z}\Big)(p^{2}_{x}-p^{2}_{y}+p^{2}_{z})\Big)\Bigg) \Bigg], \nonumber \\
A_{T\lambda=0}^{(2^{+},L=3)0}(|\vec{q}\,|)=&
-i\sqrt{\frac{3}{4}}
\sqrt{2m_{I}2E_{F}(-\vec{q}\,)}\int
d^{3}p\sqrt{\frac{\hat{E}_{f'_{1}}\hat{E}_{f_{1}}}{4E_{f'_{1}}E_{f_{1}}}}\frac{1
}{4\pi p}
\left(
\hat{\phi}_{f'_{1}f_{2}}^{(M(2^{+},L=3)}( p)\right)^{\ast}
\nonumber\\
&\times\hat{\phi}_{f_{1}f_{2}}^{(M(0^{-}))}
(|\vec{p}-\frac{m_{f_{2}}}{m_{f'_{1}}+m_{f_{2}}}|\vec q\,|\vec{k}\,|)
\Bigg[\left(\frac{5p^{2}_{z}}{p^{2}}-1\right)
\left(\frac{p^{2}_{x}+p^{2}_{y}}{\hat{E}_{f_{1}}}
+\frac{p^{2}_{x}+p^{2}_{y}}{\hat{E}_{f'_{1}}}\right) \nonumber \\
&
-\frac{p_{z}}{p}\left(\frac{5p^{2}_{z}}{p^{2}}
-3\right)\Bigg(\frac{\frac{m_{f_{2}}}{m_{f'_{1}}+m_{f_{2}}}|\vec{q}\,|-p_{z}}{
\hat{E}_{f_{1}}}
-\frac{\frac{m_{f'_{1}}}{m_{f'_{1}}+m_{f_{2}}}|\vec{q}\,|+p_{z}}{\hat{E}_{f'_{1}
}}\Bigg)\Bigg], \nonumber \\
A_{\lambda=0}^{(2^{+},L=3)3}(|\vec{q}\,|)=&
-\frac{i}{2}\sqrt{2m_{I}2E_{F}(-\vec{q}\,)}\int
d^{3}p\sqrt{\frac{\hat{E}_{f'_{1}}\hat{E}_{f_{1}}}{4E_{f'_{1}}E_{f_{1}}}}
\frac{1}{4\pi p}\left(\hat{\phi}_{f'_{1}f_{2}}^{(M(2^{+},L=3))}(
p)\right)^{\ast} \nonumber\\
&
\times\hat{\phi}_{f_{1}f_{2}}^{(M(0^{-}))}(|\vec{p}-\frac{m_{f_{2}}}{
m_{f'_{1}}+m_{f_{2}}}|\vec q\,|\vec{k}\,|)\left[
(p^{2}_{x}+p^{2}_{y})\left(\frac{5p^{2}_{z}}{p^{2}}-1\right)\left(\frac{\frac{m_
{f_{2}}-m_{f_{1}'}}{m_{f'_{1}}+m_{f_{2}}}|\vec{q}\,|-2p_{z}}{\hat{E}_{f_{1}}\hat
{E}_{f'_{1}}}\right) \right. \nonumber \\
&
\left.
-p_{z}\left(\frac{5p^{2}_{z}}{p^{2}}-3\right)\left(1-\frac{p^{2}_{x}+p^{2}_{y}
-\left(-\frac{m_{f'_{1}}}{m_{f'_{1}}+m_{f_{2}}}|\vec{q}\,|-p_{z}
\right)\left(\frac{m_{f_{2}}}{m_{f'_{1}}+m_{f_{2}}}|\vec{q}\,|-p_{z}\right)}{
\hat{E}_{f_{1}}\hat{E}_{f'_{1}}}\right)\right], \nonumber \\
A_{\lambda=+1}^{(2^{+},L=3)1}(|\vec{q}\,|)=&
-\frac{i}{\sqrt{8}}\sqrt{2m_{I}2E_{F}(-\vec{q}\,)}\int
d^{3}p
\sqrt{\frac{\hat{E}_{f'_{1}}\hat{E}_{f_{1}}}{4E_{f'_{1}}E_{f_{1}}}}\frac{1}{4\pi
p}\left(\hat{\phi}_{f'_{1}f_{2}}^{(M(2^{+},L=3))}(p)\right)^{\ast}
 \nonumber\\
&
\times\hat{\phi}_{f_{1}f_{2}}^{(M(0^{-}))}(|\vec{p}-\frac{m_{f_{2}}}{m_{f'_{1}}
+m_{f_
{2}}}|\vec q\,|\vec{k}\,|)\Bigg[
3p_{z}
\nonumber \\
&+3p_{z}\frac{p^{2}_{x}-p^{2}_{y}-\left(-\frac{m_{f'_{1}}}{m_{f'_{1}}+m_{
f_{2}}}|\vec{q}\,|-p_{z}\right)\left(\frac{m_{f_{2}}}{m_{f'_{1}}+m_{f_{2}}}|\vec
{q}\,|-p_{z}\right)}{\hat{E}_{f_{1}}\hat{E}_{f'_{1}}} \nonumber \\
&
+5p_{z}\left(\frac{p^{2}_{x}}{p^{2}}+\frac{p^{2}_{y}}{p^{2}}-\frac{p^{2}_{z}}{p^
{2}}\right)\left(1+\frac{p^{2}_{x}-p^{2}_{y}-\left(-\frac{m_{f'_{1}}}{m_{f'_{1}}
+m_{f_{2}}}|\vec{q}\,|-p_{z}\right)\left(\frac{m_{f_{2}}}{m_{f'_{1}}+m_{f_{2}}}
|\vec{q}\,|-p_{z}\right)}{\hat{E}_{f_{1}}\hat{E}_{f'_{1}}}\right) \nonumber \\
&
-2p^{2}_{x}\left(\frac{5p^{2}_{z}}{p^{2}}-1\right)\left(\frac{\frac{m_{f_{2}}-m_
{f_{1}'}}{m_{f'_{1}}+m_{f_{2}}}|\vec{q}\,|-2p_{z}}{\hat{E}_{f_{1}}\hat{E}_{f'_{1
}}}\right)+20\frac{p_{z}p^{2}_{x}p^{2}_{y}}{\hat{E}_{f_{1}}\hat{E}_{f'_{1}}p^{2}
} \Bigg]. \nonumber \\
\end{align}

\end{enumerate}

\end{widetext}

\end{document}